\documentclass[11pt]{article}

\usepackage{mathtools}

\usepackage[normalem]{ulem}

\usepackage[all]{xy}

\usepackage{amsmath}
\usepackage{amsfonts}
\usepackage{appendix}
\usepackage{verbatim}

\usepackage{amsmath}
\usepackage{amssymb}
\usepackage{latexsym}
\usepackage{graphicx}
\usepackage[english]{babel}
\usepackage[font={small}]{caption}
\usepackage{mathtools}
\usepackage{geometry}                
\usepackage{epstopdf}
\usepackage{tikz-cd}
\usetikzlibrary{cd}
\usepackage{amsfonts}
\usepackage{amssymb}
\usepackage{latexsym}
\usepackage{graphicx}
\usepackage[english]{babel}
\usepackage{tikz}

\usepackage{cite}

\begin{document}
\input epsf

\def\p{\partial}
\def\h{{1\over 2}}
\def\be{\begin{equation}}
\def\bea{\begin{eqnarray}}
\def\ee{\end{equation}}
\def\eea{\end{eqnarray}}
\def\d{\partial}
\def\la{\lambda}
\def\eps{\epsilon}
\def\b{\bigskip}
\def\m{\medskip}

\newcommand{\newsection}[1]{\section{#1} \setcounter{equation}{0}}

\def\q{\quad}

\def\h{{1\over 2}}
\def\t{\tilde}
\def\r{\rightarrow}
\def\nn{\nonumber\\}

\let\p=\partial

\newcommand\blfootnote[1]{%
  \begingroup
  \renewcommand\thefootnote{}\footnote{#1}%
  \addtocounter{footnote}{-1}%
  \endgroup
}

\begin{flushright}
\end{flushright}
\vspace{20mm}
\begin{center}
{\LARGE Dynamical evolution in the  D1D5 CFT}
\\
\vspace{18mm}
{\bf   Bin Guo$^1$\blfootnote{$^{1}$bin.guo@ipht.fr}, Samir D. Mathur$^2$\blfootnote{$^{2}$mathur.16@osu.edu}
\\}
\vspace{10mm}
${}^1$Institut de Physique Th\'eorique,\\
	Universit\'e Paris-Saclay,
	CNRS, CEA, \\ Gif-sur-Yvette, 91191, France  \\ \vspace{4mm}

${}^2$Department of Physics,\\ The Ohio State University,\\ Columbus,
OH 43210, USA\\

\vspace{8mm}
\end{center}

\vspace{4mm}

\thispagestyle{empty}
\begin{abstract}

 It is interesting to ask: how does the radial space direction emerge from the CFT in gauge-gravity duality?
In this context we resolve a long-standing puzzle with the gravity duals of two classes of states in the D1D5 CFT. For each class the CFT states are in the untwisted sector, suggesting that the energy gap should be $1/R_y$ where $R_y$ is the radius of the circle on which the D1D5 CFT is compactified. For one class of states, the gravity dual indeed has exactly this gap, while for the other class, the gravity dual has a very deep throat, leading to an energy gap much smaller than $1/R_y$. We resolve this puzzle by showing that for the latter class of states, perturbing the CFT off its free point  leads to the formation of a band structure in the CFT. We also explain why such a band structure does not arise for the first class of states. 
Thus for the case where a deep throat emerges in the gravity description, the dynamics of falling down this throat is described in the CFT as a sequential `hopping' between states all of which have the same energy at the free point; this hopping amplitude converts an integer spaced spectrum into a closely spaced band of energy levels.

\vspace{3mm}

\end{abstract}
\newpage

\setcounter{page}{1}

\numberwithin{equation}{section} 

\tableofcontents

\newpage

\section{Introduction}

The bound state of D1 and D5 branes gives a very useful example of AdS/CFT duality \cite{adscft1,adscft2,adscft3}.  The CFT is 1+1 dimensional, and is conjectured to have an `orbifold point' in its moduli space where it is weakly coupled and relatively easy to study \cite{Vafa:1995bm,Dijkgraaf:1998gf,orbifold2,Larsen:1999uk,Arutyunov:1997gt,Arutyunov:1997gi,Jevicki:1998bm,David:2002wn}.  Semiclassical gravity  emerges in a very different region of moduli space, where the CFT is strongly coupled. Nevertheless many quantities of interest are found to agree between the orbifold point and the semiclassical gravity domain; examples are the extremal and near extremal entropies of black holes, and the emission spectrum from the near extremal holes \cite{sen,sv,cm,dmcompare,maldastrom}.

The goal of this paper is to study a situation where the dynamics of the CFT at the orbifold point differs strongly from the dynamics at the semiclassical gravity point. This difference in behavior is of course not a contradiction, since the two theories are at different couplings. We examine the deformation of the CFT off the orbifold point. We find that this deformation modifies the dynamics of the CFT so that we can understand the behavior found at the supergravity point. We believe that this analysis provides an interesting illustration for how AdS/CFT duality leads to the emergence of spacetime from a large $N$ field theory. 

\subsection{The nature of 2-charge extremal states}\label{sec nature}

We will consider the compactification 
\be
{\mathcal M}_{9,1}\r {\mathcal M}_{4,1}\times S^1\times T^4
\label{compact 0}
\ee
 though similar results should hold for ${\mathcal M}_{9,1}\r {\mathcal M}_{4,1}\times S^1\times K3$. The D1D5 CFT describes the bound state of D1 branes wrapped on $S^1$ and D5 branes wrapped on $S^1\times T^4$. Let the bound state have $n_1$ D1 branes and $n_5$ D5 branes.  The ground state of this theory is highly degenerate, with a degeneracy  $\text{Exp}[2\pi\sqrt{2}\sqrt{n_1n_5}]$. We can explicitly describe all these states at the orbifold point, and also construct the states at strong coupling where they become `fuzzballs' \cite{lm4,lmm,skenderis,threecharge1,threecharge2,threecharge3,Bena:2016ypk,Heidmann:2019zws,Bena:2022sge,Ganchev:2021iwy}. We are interested in low energy dynamics around these ground states; in particular, 1-particle excitations of the gravity solutions, and the  excitations of the corresponding CFT state. 

To understand the issue, we have to look at the structure of the microstates in the CFT and in the gravity theory:

\b

(i) The CFT at the orbifold point is described by 
$N\equiv n_1n_5$
 copies of a free $c=6$ CFT. The $S_N$ orbifolding symmetrizes  these $N$ copies, and in the process also allows twist sectors where $k$ copies of the CFT are linked together to give a single copy of the CFT on a circle of length $2\pi k R$.  We call each such set of linked copies a `component string'. We consider the CFT in the Ramond (R) sector, which is the natural description for the D1D5 bound state sitting in asymptotically flat spacetime. Each component string then has a 16-fold degeneracy: there are 4 ground states in the left sector and 4 in the right sector.  We call the left ground states $|0^+\rangle^{[k]}_R, |0^-\rangle^{[k]}_R, |0\rangle^{[k]}_R, |\t 0\rangle^{[k]}_R$, and  the right ground states $|\bar 0^+\rangle^{[k]}_R, |\bar 0^-\rangle^{[k]}_R, |\bar 0\rangle^{[k]}_R, |\bar{\t 0}\rangle^{[k]}_R$; here $[k]$ labels the twist sector of the component string. 

\b

(ii) To construct the microstates in the gravity theory we first use S,T dualities to map the D1D5 bound state to an NS1P bound state: we get $n_5$ units of winding for the elementary string NS1, and $n_1$ units of momentum charge P along this NS1. The NS1P bound state is just given by a single NS1 with winding $n_5$, carrying $n_1$ units of momentum in the form of travelling waves along this NS1. This momentum is carried by bosonic and fermionic oscillators $\alpha^i_{-{k/ n_5}}$ and $d^i_{-{k/ n_5}}$. Here $k=1, 2, \dots$, and $i=1, \dots 8$ gives the polarization of the oscillator in the 8 spatial directions transverse to the NS1. The gravity solution for any profile of a vibrating string can be explicitly constructed, and transforming this solution back to the D1D5 duality frame using S,T dualities one gets the gravity solution describing the microstate.

\b

The CFT states above are at the orbifold point, and the gravity states are at a different point in moduli space. Nevertheless, matching various quantum numbers allows us to make a heuristic map between the two constructions:

\b

(a) Consider an oscillator $\alpha^i_{-{k/ n_5}}$ or $d^i_{-{k/ n_5}}$ in the construction (ii). This maps to a component string with twist $k$ in the orbifold theory (i).

\b

(b) The polarization $i$ of an oscillator like  $\alpha^i_{-{k/ n_5}}$ or $d^i_{-{k/ n_5}}$ maps to choice of ground state for the component string: there are 8 choices of $i$ for the bosonic oscillators and 8 for the fermionic oscillators, and these map to the 16 possible ground states for the $k$-twisted component string.

\b

\subsection{The puzzle}

To understand the puzzle, look at the 8 different polarizations $i$ of the oscillators $\alpha^i_{-{k/ n_5}}$ in the NS1P duality frame. The spacetime is compactified as (\ref{compact 0}). The NS1 is wrapped along the $S^1$. The 8 transverse directions fall into two groups (a) $i=1,2,3,4$, corresponding to the directions along the $T^4$, and (b)  $i=5,6,7,8$, corresponding to the 4 noncompact directions. These noncompact directions can be written in terms of a radial coordinate $r$ which becomes the radial coordinate of $AdS_3$ in the D1D5 duality frame, and an angular 3-sphere which becomes the $S^3$ in the D1D5 frame.

The vibrations carrying the momentum P give the NS1 a transverse vibration profile $F^i(t-y)$, where  $y$ is the coordinate along the $S^1$. The $F^i$ are periodic under $y\r y+2\pi n_5 R_y$, where $R_y$ is the radius of the $S^1$ and we have noted that the NS1 winds  $n_5$ times around this $S^1$ before closing. 

\begin{figure}[htbp]
\begin{center}
\includegraphics[scale=0.32]{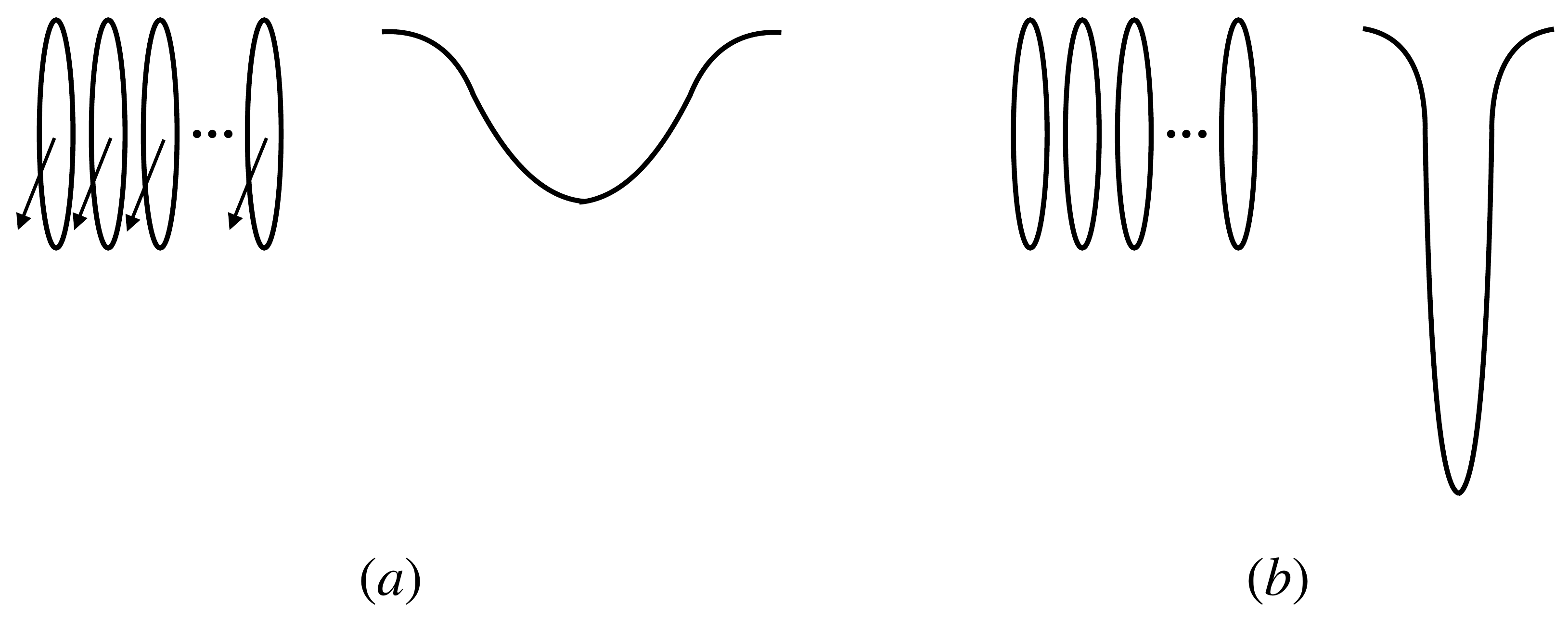}\hskip20pt
\end{center}
\caption{(a) Singly wound strings $\left ( |0^+\rangle^{[1]}_R |\bar 0^+\rangle^{[1]}_R \right )^{N}$ and their gravity dual. (b) Singly wound strings $\left ( |0\rangle^{[1]}_R |\bar 0\rangle^{[1]}_R \right )^{N}$ and their gravity dual.}
\label{fig1}
\end{figure}

Without this vibration profile, the charges NS1 and P would generate what we call the `naive' geometry (in the string frame)
\be
ds^2_{string}=-(1+{Q_1\over r^2})^{-1}[-dt^2+dy^2] + (1+{Q_5\over r^2})[dr^2+r^2d\Omega_3^2] + \sum_{i=6}^9 dx^idx^i
\label{naiveg}
\ee
The vibrations in the noncompact directions cause the NS1 to move from $r=0$ to $r>0$, and thus alters the gravity solution upto some radius $r\sim \bar r$. Thus the $AdS_3$ geometry is a good approximation for $r\gtrsim \bar r$; for $r\lesssim \bar r$ we get a `fuzzball' whose structure reflects the choice of microstate. But vibrations along the compact $T^4$ directions do not move the NS1 in the radial direction to $r>0$; they move the string in the $T^4$, and since this $T^4$ is small and compact, the strands of the NS1 never suffer a large displacement under the vibrations in the torus directions. This situation is depicted in fig.\ref{fig1}: when the vibrations are in the noncompact directions $i=5,6,7,8$, the microstates geometry have a short throat terminating in a fuzzball `cap', and when the vibrations are in the compact $T^4$ directions $i=1,2,3,4$, the microstates geometry have a throat that is deep throat reaching very close to $r=0$ before quantum effects produce a cap.

Let us state this more concretely with the help of examples, since our puzzle derives from this difference between the compact and noncompact vibration directions:

\b

(1) Consider the following simple example of NS1P vibrations in the noncompact directions. Writing $v=t-y$, we take the vibration profile  
\be
F^5=a \sin ({k  v\over n_5 R_y}), ~~~F^6=a \cos ({k v\over n_5 R_y})
\label{vib1}
\ee
with other the $F^i=0$.
Thus the NS1 vibration profile executes $k$ turns of a uniform helix. The gravity solution generated by this NS1P profile, and the corresponding solution in the D1D5 frame, were found in \cite{lm4}. Such a D1D5 solution is depicted  in fig.\ref{fig1}(a). 

The corresponding state of the orbifold CFT is
\be
\left ( |0^+\rangle^{[k]}_R |\bar 0^+\rangle^{[k]}_R \right )^{N\over k}
\label{cft1}
\ee
i.e., there are $N/k$ component strings, each with twist $k$, and each being in the Ramond ground state $|0^+\rangle^{[k]}_R |\bar 0^+\rangle^{[k]}_R$. 

Now consider the spectrum of excitations in both the gravity and CFT descriptions. In the gravity  description, we find massless scalars satisfying $\square \phi =0$. Solving this equation in the D1D5 geometry, we find solutions
\be
\phi_n=\t \phi_n(r) e^{-i {2 n\over kR_y} t}
\ee
so that the energy gap is
\be
\Delta E_{grav} = {2\over kR_y}
\label{egrav}
\ee
In the CFT description, the scalar field excitation is described by one left oscillator and one right oscillator like
\be
\alpha_{-+,-\frac{n}{k}}\bar \alpha_{-+,-\frac{n}{k}}
\ee
where the charge indices will be explained later in section \ref{sec sym}.
These oscillators should not be confused with the bosonic oscillators $\alpha^i_{-{k/ n_5}}$ describing string profiles in section \ref{sec nature}.
This excitation has energy ${2n\over k R_y}$, so the energy gap is again
\be
\Delta E_{CFT} = {2\over kR_y}
\label{ecft}
\ee
This agreement between (\ref{egrav}) and (\ref{ecft}) provided an interesting confirmation of the fuzzball structure of gravity microstates. 

\b

(2) Now consider  NS1P vibrations in the compact directions. Take the vibration profile  
\be
F^1=a \sin ({k  v\over n_5 R_y}), ~~~F^2=a \cos ({k v\over n_5 R_y})
\label{vib2}
\ee
with other the $F^i=0$.
While this NS1 vibration profile appears to executes $k$ turns of a uniform helix with radius $a$, the $T^4$ directions  are compactified with a small (string scale) radius, so the NS1 strands move are displaced by only a small distance which we will call $\epsilon$; this number depends on the size of the $T^4$ and quantum corrections to the classical string profile.  Dualizing to the D1D5 frame, the geometry is well approximated by (\ref{naiveg}) down to some small radius $r=\epsilon$.
Such a D1D5 solution is depicted  in fig.\ref{fig1}(b). In this deep geometry, the energy gap for excitations of the scalar field $\square \phi=0$ is very small 
\be
\Delta E_{grav}\ll {2\over kR_y} 
\label{egrav2}
\ee

The corresponding state of the orbifold CFT is 
\be
\left ( |0\rangle^{[k]}_R |\bar 0\rangle^{[k]}_R \right )^{N\over k}
\label{cft2}
\ee
 Note that the vacuua in (\ref{cft2}) are different from those in (\ref{cft1}); this difference will be at the heart of our analysis. The vacuua in (\ref{cft1}) carry charges under the $su(2)\times su(2)\approx so(4)$ that gives the symmetry group of $S^3$ in (\ref{compact}); these charges describe the fact that the vibrations (\ref{vib1}) in the noncompact directions $x_6,x_7$ break the rotation symmetry of this $S^3$. The vibrations (\ref{vib2}) in the torus directions $x_1,x_2$ do not break this symmetry, and so the vacuua in (\ref{cft2}) do not carry charges under this $su(2)\times su(2)\approx so(4)$. 

In the CFT state (\ref{cft2}), the scalar field excitations at the orbifold point are described by one left oscillator and one right oscillator like
\be
\alpha_{-+,-\frac{n}{k}}\bar \alpha_{-+,-\frac{n}{k}}
\ee
This excitation again has energy ${2n\over k R_y}$, so the energy gap is
\be
\Delta E_{CFT} = {2\over kR_y}
\label{ecft2}
\ee
This disagreement between (\ref{egrav2}) and (\ref{ecft2}) is the puzzle that we are interested in. Note that the disagreement is not a contradiction, since (\ref{egrav2}) and (\ref{ecft2}) are at different points in moduli space. Rather the issue is: why do we get such good agreement between these different points in moduli space for the  CFT states (\ref{cft1}) but get such a large  disagreement for the states (\ref{cft2}).  Thus our goal is to understand how the deformation off the orbifold point changes the energy gaps in the CFT from the  large values (\ref{ecft2}) to the small values (\ref{egrav2}).

\subsection{Results}\label{sec results}

We will study the perturbation theory around the obifold point generated by the deformation operator $D$. This operator consists of a twist $\sigma_2$ which twists together two copies of the $c=6$ CFT, together with left and right supercharges $G, \bar G$. Both these aspects of $D$ will be relevant to us.

The following is a schematic description of our computations and results. Consider a Ramond ground state where all copies of the CFT are singly wound, and have a state corresponding to a $T^4$ mode
\be\label{ls}
\left ( |0\rangle^{[1]}_R |\bar 0\rangle^{[1]}_R \right )^{N} = \left(|0\rangle^{[1]}_R |\bar 0\rangle^{[1]}_R\right)\left(|0\rangle^{[1]}_R |\bar 0\rangle^{[1]}_R\right)
 {\dots}  \left(|0\rangle^{[1]}_R |\bar 0\rangle^{[1]}_R  \right)
\ee
The superscript $[1]$ on a state denotes that the state describes a component string with winding $k=1$; i.e.,the component string is  in the untwisted sector.
On one of the copies we add energy in the form of one left and one right moving excitation, in their lowest allowed harmonic.
\be\label{psi 1}
|\psi_1\rangle=\left(\alpha_{-+,-1}^{[1]}\bar\alpha_{-+,-1}^{[1]} |0\rangle^{[1]}_R |\bar 0\rangle^{[1]}_R \right ) \left ( |0\rangle^{[1]}_R |\bar 0\rangle^{[1]}_R \right )^{N-1}
\ee
where the left and right moving excitations $\alpha_{-+,-1}^{[1]}, \bar\alpha_{-+,-1}^{[1]}$ are placed on one of the component strings in the state $|0\rangle^{[1]}_R |\bar 0\rangle^{[1]}_R$, say the first copy. At the orbifold point this state would be an eigenstate of the Hamiltonian. But away from the orbifold point, we have the action of the deformation operator which can twist copy 1 with another copy, say copy 2, to yield a doubly wound component string. Then the state (\ref{psi 1}) can evolve into a state like
\be
|\psi_2\rangle=\left((\alpha_{-+,-\h}^{[2]})^2(\bar\alpha_{-+,-\h}^{[2]})^2 |0\rangle^{[2]}_R |\bar 0\rangle^{[2]}_R \right)\left ( |0\rangle^{[1]}_R |\bar 0\rangle^{[1]}_R \right )^{N-2}
\ee
where the two left and two right moving excitations $(\alpha_{-+,-\h}^{[2]})^2(\bar\alpha_{-+,-\h}^{[2]})^2$ are placed on the doubly wound string, while all the other component strings remain singly wound.

A second insertion of the deformation operator can join another singly wound copy to the doubly wound copy in $|\psi_2\rangle$, generating a component string with winding $3$, on which the excitation energy is carried by 3 left and 3 right excitations with energy $1/3$ each. In this manner we get a set of states $|\psi_k\rangle$, all with the same total energy, and the deformation operator $D$ connects each such state to the next state in the sequence. 
\be\label{psi k}
|\psi_k\rangle=\left((\alpha_{-+,-\frac{1}{k}}^{[k]})^k(\bar\alpha_{-+,-\frac{1}{k}}^{[k]})^k |0\rangle^{[k]}_R |\bar 0\rangle^{[k]}_R \right)\left ( |0\rangle^{[1]}_R |\bar 0\rangle^{[1]}_R \right )^{N-k}
\ee

\begin{figure}[htbp]
\begin{center}
\includegraphics[scale=0.4]{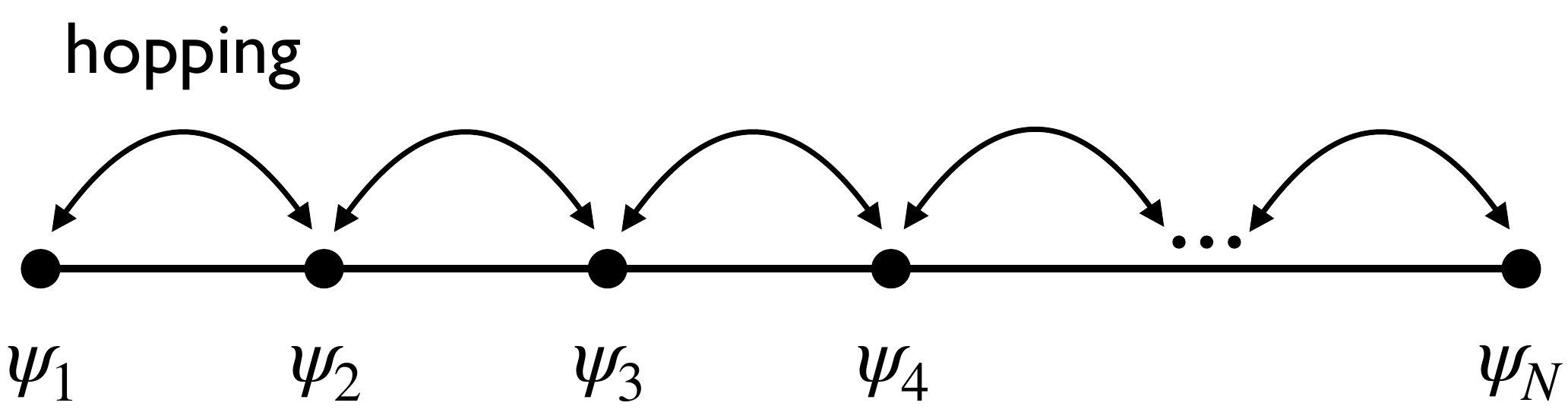}\hskip20pt
\end{center}
\caption{A 1-dimensional lattice with a hopping interaction between nearby points.}
\label{fig2}
\end{figure}

Now imagine these states to be represented by a 1-dimensional lattice of points as depicted in fig.\ref{fig2}, with $|\psi_1\rangle$ the first point, $|\psi_2\rangle$ the second point and so on. (We will see later that of dimension higher than $1$ emerges,  but since we are ignoring spin indices etc., this crude model will suffice to explain the essence of the physics.) The action of $D$ is then a hopping term on this lattice, enabling hopping from any one point to its immediate neighbor. We compute this hopping amplitude $A(k)$ from $|\psi_k\rangle$ to $|\psi_{k+1}\rangle$ as a function of $k$ in the limit $k\rightarrow \infty$. For states of the form $(\ref{psi k})$ where the singly wound copies are of type $|0\rangle^{[1]}_R |\bar 0\rangle^{[1]}_R, |0\rangle^{[1]}_R |\bar{ \t 0}\rangle^{[1]}_R, |\t 0\rangle^{[1]}_R |\bar 0\rangle^{[1]}_R, |\t 0\rangle^{[1]}_R |\bar{ \t 0}\rangle^{[1]}_R$, we find that
\be
A(k)\sim 1
\label{el}
\ee
i.e., the amplitude of hopping does not grow or decay with $k$. 

\bigskip

The situation is now similar to that of a crystal,  where we have a chain of atoms. An electron can reside with the same energy at any of the atoms, and there is an interaction which can hop the electron from one atom to a neighboring atom. The energy levels of the electron in such a crystal form a {\it band}, with the level spacing behaving as $\Delta E \sim 1/N$ where $N$ are the number of atoms in the chain. Similarly, in our problem we get a band, with spacing that will depend on the number of states denoted by dots in fig.\ref{fig2}; this number is equal to the maximal value of the twist $k$, which is $N=n_1n_5$. Thus the energy gaps become very small, and this is in accordance with (\ref{egrav2}).

We find this band formation interesting as it gives an illustration of how spacetime emerges from the CFT: for the states (\ref{cft2}), the dual gravity solutions have a deep throat, and the geometry of this deep throat reflects band formation in the CFT, an effect that arises only after deformation off the `free' orbifold point is considered.

One could now ask why a similar band structure does not arise for the CFT states (\ref{cft1}). Such a state has the form 
\be\label{sixp}
|\psi_1\rangle=\left(\alpha_{-+,-1}^{[1]}\bar\alpha_{-+,-1}^{[1]} |0^+\rangle^{[1]}_R |\bar 0^+\rangle^{[1]}_R \right ) \left ( |0^+\rangle^{[1]}_R |\bar 0^+\rangle^{[1]}_R \right )^{N-1}
\ee
We can again try to look at the process where  more and more singly wound copies join to make longer strings. But this time each singly wound copy is in  the state $ |0^+\rangle^{[1]}_R |\bar 0^+\rangle^{[1]}_R$, so it carries charge $\h$ under each of the groups $SU(2)_L, SU(2)_R$. The $k$-wound component string multiwound copy must therefore carry charges $\sim k/2$ for each of these groups. These charges must be carried by fermionic excitations, since the bosonic excitations do not carry $SU(2)_L, SU(2)_R$ charges.  
 Due to the Pauli exclusion principle, these fermions cannot all be in low energy states. They must  at least have energies $1/k, 2/k, \dots, k/k$.
So the total energy of these fermions rises linearly with $k$.
Since the excitation energy that we started with in (\ref{sixp}) is a given finite quantity, we cannot keep joining more and more singly wound copies to get a larger and larger $k$.  
This explains how an agreement between (\ref{ecft}) and (\ref{egrav}) is possible.

There have been many works that study conformal perturbation theory in the context of the D1D5 CFT, for example \cite{gn,Avery:2010er,Avery:2010hs,Pakman1,Pakman2,Pakman3,Burrington:2012yq,Burrington:2017jhh,c6,c5,c4,c3,c2,c1,Hampton:2019csz,gz,hmz,Keller:2019suk,Keller:2019yrr,Benjamin:2021zkn,gm1,gm2,gm3,gh,gh1,gh2,l1,l2,l3,l4,l5,l6,Dei:2019iym,Apolo:2022fya}. 
For more computations in  conformal perturbation theory in two and higher dimensional CFTs see, e.g.  \cite{kadanoff,Dijkgraaf:1987jt,Cardy:1987vr,Kutasov:1988xb,Eberle:2001jq,Gaberdiel:2008fn,Berenstein:2014cia,Berenstein:2016avf}.
There is also recent progress in deriving the tensionless string limit of the AdS/CFT correspondence using the orbifold D1D5 CFT \cite{t1,t2,t3,t4,t5,t6,t7,t8}.

\section{The D1D5 CFT}

In this section, we briefly summarize some properties of the D1D5 CFT at the orbifold point and the deformation operator that we will use to perturb away from the orbifold point. 
For more details, see \cite{Avery:2010er,Avery:2010hs}.

Consider the compactification of type IIB string theory
\be
M_{9,1}\rightarrow M_{4,1}\times S^1\times T^4.
\label{compact}
\ee
Wrap $n_1$ D1 branes on $S^1$, and $n_5$ D5 branes on $S^1\times
T^4$. We think of the $S^1$ as being large compared to the $T^4$, so
that at low energies we look for excitations only in the direction
$S^1$. At this low energy limit, the bound state of these branes is described by a conformal field theory (CFT) on the circle $S^1$.

It has been conjectured that in the moduli space of couplings there exists a point called the `orbifold point' where the CFT is particularly simple. 
At this orbifold point the CFT is
a 1+1 dimensional sigma model. We will work in the Euclidized theory, where
the base space is a cylinder with the coordinates 
\be
\tau, \sigma: ~~~0\le \sigma<2\pi, ~~~-\infty<\tau<\infty
\ee
where $\sigma$ parameterizes the circle $S^1$.
The target space of the sigma model is the `symmetrized product' of
$n_1 n_5$ copies of $T^4$,
\be
(T^4)^{n_1 n_5}/S_{n_1 n_5},
\ee
where each copy of $T^4$ gives 4 bosonic excitations $X^1, X^2, X^3,
X^4$. It also gives 4 fermionic excitations, which we call $\psi^1,
\psi^2, \psi^3, \psi^4$ for the left movers, and $\bar\psi^1,
\bar\psi^2,\bar\psi^3,\bar\psi^4$ for the right movers. The fermions can be
antiperiodic or periodic around the circle $S^1$. If they are
antiperiodic we are in the Neveu-Schwarz (NS) sector, and
if they are periodic we are in the Ramond (R)
sector. The central charge of the theory with fields
$X^i, \psi^i, ~i=1\dots 4$ is $c=6$. 
The total central charge of the entire system is
\be\label{N}
c=6 n_1 n_5\equiv 6N
\ee

\subsection{Symmetries of the CFT}\label{sec sym}

The D1D5 CFT has $\mathcal{N}=(4,4)$ supersymmetry, which means that we have
$\mathcal{N}=4$ supersymmetry in both the left and right moving
sectors.
This leads to a superconformal ${\cal N}=4$ symmetry generated by operators $L_{n}, G^\pm_{\pm,r},
J^a_n$ for the left movers and $\bar L_{n}, \bar G^\pm_{\pm,r}, \bar
J^a_n$ for the right movers. The full symmetry is actually larger, which is the contracted large $\mathcal{N}=4$ superconformal symmetry \cite{Schwimmer:1986mf,Sevrin:1988ew}. The generators and commutators of the algebra are given in Appendix~\ref{conventions}. 

The internal R symmetry group for each ${\cal N} = 4$ algebra is
$SU(2)$, so there is
a global symmetry group $SU(2)_L\times SU(2)_R$.  The
quantum numbers in these two $SU(2)$ groups are denoted by
\be
SU(2)_L: ~(j, m);~~~~~~~SU(2)_R: ~ (\bar j, \bar m).
\ee
This symmetry arises from the
rotational symmetry in the four noncompact space directions of $M_{4,1}$:  we have $SO(4)_E\simeq SU(2)_L\times SU(2)_R$.
Here the subscript $E$ stands for `external', which means that these
rotations are in the noncompact directions.  We have another $SO(4)$ symmetry in the four compact directions
of the $T^4$. We call this symmetry $SO(4)_I$, where $I$ stands for
`internal'. This symmetry is broken by the compactification of the
$T^4$, but it still provides a useful organizing
principle at the orbifold point. We have $SO(4)_I\simeq SU(2)_1\times SU(2)_2$.
We use spinor indices $\alpha, \bar\alpha$ for $SU(2)_L$ and $SU(2)_R$
respectively. We use spinor indices $A, \dot A$ for $SU(2)_1$ and
$SU(2)_2$ respectively.

The 4 real fermions of the left sector can be grouped into complex
fermions $\psi^{\alpha A}$. The right moving fermions have indices $\bar{\psi}^{\bar\alpha  A}$. The bosons $X^i$ are a vector in the
$T^4$. This vector can be decomposed into the $(\h, \h)$  representation of $SU(2)_1\times SU(2)_2$, which gives  scalars $X_{A\dot A}$.

\subsection{Deformation of the CFT}

To move away from the orbifold point, we add a deformation operator $D$ to the Lagrangian
\be\label{defor S}
S\r S+\lambda \int d^2 z D(z, \bar z)
\ee
where $D$ has conformal dimensions $(h, \bar h)=(1,1)$. A choice of $D$ which is a singlet under all the symmetries at the orbifold point is
\be\label{D 1/4}
D=\frac{1}{4}\epsilon^{\dot A\dot B}\epsilon_{\alpha\beta}\epsilon_{\bar\alpha \bar\beta} G^{\alpha}_{\dot A, -\h} \bar G^{\bar \alpha}_{\dot B, -\h} \sigma^{\beta \bar\beta} 
=  -\, G^{-}_{+, -\h} \bar G^{-}_{-, -\h} \sigma^{+ +} + G^{-}_{-, -\h} \bar G^{-}_{+, -\h} \sigma^{+ +}
\ee
where $ G$ and $\bar G$ are the left and right moving supercharge operators at the orbifold point.
The operator $\sigma^{\beta\bar\beta}$ is a twist operator  in the orbifold theory which can twist together any two copies of the $c=6$ CFT. We have
\bea
\sigma^{\beta \bar \beta} = \sum_{i< j}\sigma^{\beta \bar \beta}_{ij}
\eea
where $\sigma^{\beta \bar \beta}_{ij}$ twists the two copies labelled by $i$ and $j$.

\subsection{The Ramond ground states}\label{sec R ground}

We will be working in the Ramond sector of the theory. Consider a $k$-twisted component string. The ground state of this component string is degenerate, on both the left and right sectors. There are $4$ Ramond ground states for the left sector
\be
|0^-\rangle_R^{[k]},~~~~|0^+\rangle_R^{[k]}=d^{++[k]}_0d^{+-[k]}_0|0^-\rangle_R^{[k]},
~~~~|0\rangle_R^{[k]}=d^{++[k]}_0|0^-\rangle_R^{[k]},~~~~
|\tilde 0\rangle_R^{[k]}=d^{+-[k]}_0|0^-\rangle_R^{[k]}
\label{qgroundstates}
\ee
and a similar $4$ Ramond ground states for the right sector. The superscript `$[k]$' corresponds to a $k$ twisted component string. Thus for any component string, there are  $16$ Ramond ground states.

\section{The effect of the deformation operator}\label{sec twist}

In this section we recall the effect of the deformation operator $D$ \cite{Avery:2010er,Avery:2010hs,c4,Guo:2022sos}. We proceed in the following steps:

\b

(A) Consider two component strings in the initial state with windings $M$ and $N$ respectively. The deformation operator $D(w_0, \bar w_0)$ contains a twist $\sigma_2$ which can join these two component strings to a component string with winding $M+N$; here $w_0$ is the position of the deformation operator on the cylinder. The initial state component strings will in general have bosonic and fermionic excitations on them; we collectively denote such excitations by $O_{i,-n_{i}}, \bar O_{i,-n_{i}}$ for the left and right movers. Our goal is to find the state on the component string of winding $M+N$. 

\b

(B) We will work for the left and right sectors separately, and combine the results at the end. The deformation operator contains left and right supercharges, which are written as contour integrals around the twist $\sigma_2$. We unwind these contours to get supercharge zero modes applied to the initial and final states. For the left sector we get
\be\label{deform left}
(G^{-}_{\dot A,-\frac{1}{2}}\sigma^{+})=\oint_{c_{\sigma}}\frac{dw}{2\pi i}G^{-}_{\dot A}(w)\,\sigma^{+}=G^{-}_{\dot A,0}\sigma^{+}-\sigma^{+}G^{-}_{\dot A,0}
\ee
In the first term, the zero mode $G^{-}_{\dot A,0}$ is placed after the twist operator, while in the second it is before the twist operator. A similar relation holds for the right supercharge.
We are then left with the problem of finding the effect of the twist $\sigma_2$ on the initial state.

\b

\begin{figure}[htbp]
\begin{center}
\includegraphics[scale=0.45]{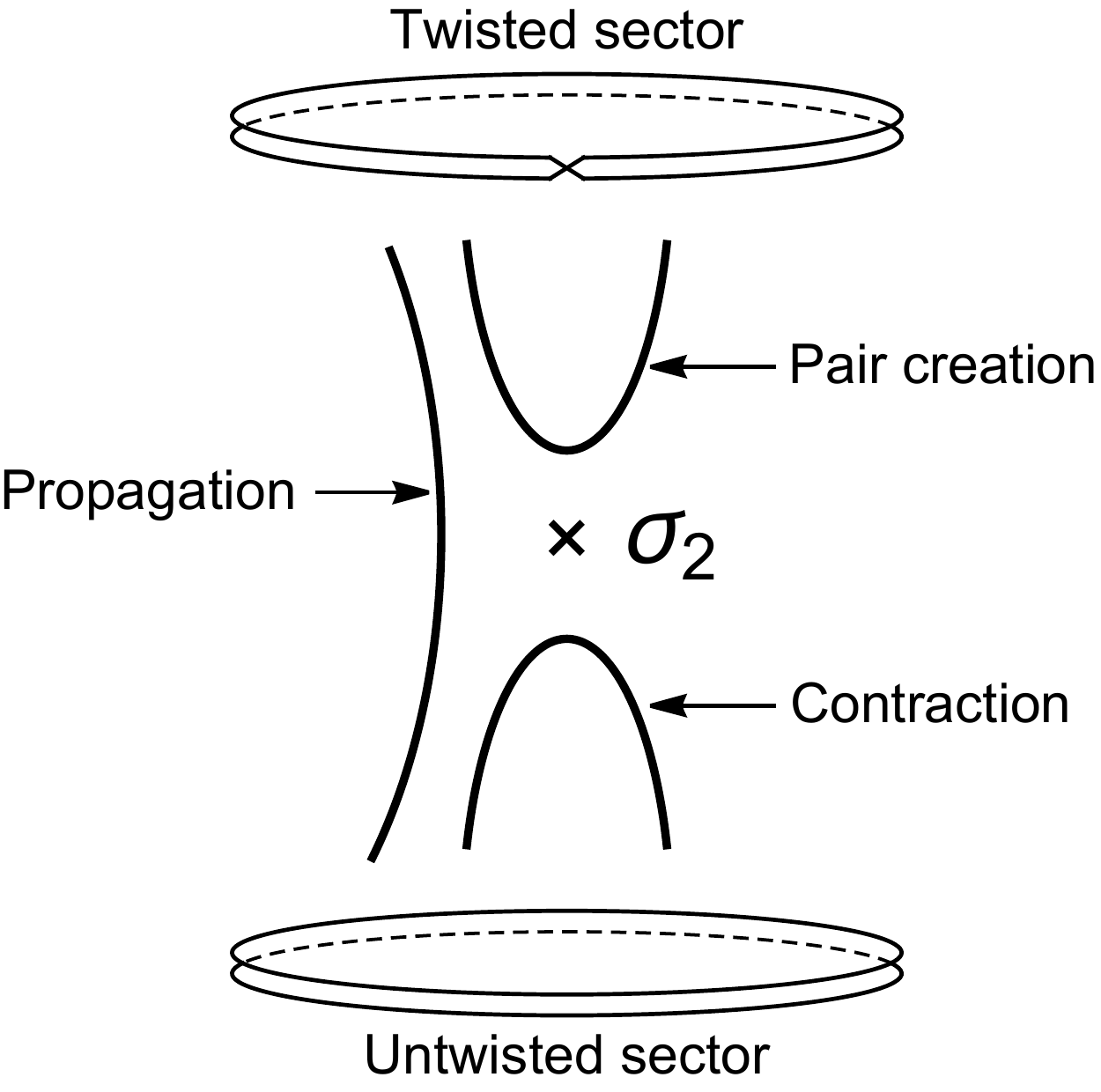}\hskip20pt
\end{center}
\caption{The `Feynman rules' to compute the effect of the twist operator}
\label{fig twist}
\end{figure}

(C) Consider the left moving state
\be\label{twist initial}
\prod_{i}O_{i,-n_{i}}|0^-\rangle_R^{[M]}|0^-\rangle_R^{[N]}
\ee
where $O_{i,-n_{i}}$ can be modes placed either on $M$ or $N$ twisted component string. We wish to find
\be
\sigma^{+}_{2}(w_{0})\Big(\prod_{i}O_{i,-n_{i}}\Big)|0^-\rangle_R^{[M]}|0^-\rangle_R^{[N]}
\ee
There are three `Feynman rules' depicted in fig.\ref{fig twist} that give the effect of the twist on the excitations $O_{i,-n_{i}}$.

\b

(i) Contraction: Two modes $O_{i,-n_{i}}$ and $O_{j,-n_{j}}$ in the initial state (\ref{twist initial}) can `Wick contract' with each other, giving a number. 
\be
C\left[O_{i,-n_{i}} O_{j,-n_{j}}\right]
\ee
Just as with the normal process of Wick contraction, one has to consider all pairs of operators that can contract: for each such possible pair, we get a term where the operators contract to a number $C$, and a term where the operators will not be contracted and will then pass through the twist as in step (ii) below. For fermionic modes, one has to include negative  signs that arise from changing the orders of the fermions to bring them next to each other to perform the contraction.

\b

(ii) Propagation: For any modes left after the contraction, we pass them through the twist to get modes  applied after the twist operator, which are modes in the $M+N$ winding sector. The modes obtained after the twist are related to the mode before the twist by the following rules:
\bea
\alpha^{[i]}_{A\dot A,-r}
\rightarrow \sigma \alpha^{[i]}_{A\dot A,-r} 
~&=&~\sum\limits_{s>0} f^{B[i]}_{rs}\,  \alpha^{[M+N]}_{A\dot A,-s} \,, \qquad i=M,N
\label{form alpha} \\
d^{+ A[i]}_{-r}
\rightarrow \sigma d^{+ A[i]}_{-r}  
~&=&~\sum\limits_{s\geq 0} f^{F[i]+}_{rs}\,  d^{+ A[M+N]}_{-s} \,, \qquad i=M,N 
\label{form d+}\\
d^{- A[i]}_{-r}
\rightarrow \sigma d^{- A[i]}_{-r}  
~&=&~\sum\limits_{s> 0} f^{F[i]-}_{rs}\,  d^{- A[M+N]}_{-s} \,, \qquad i=M,N 
\label{form d-}
\eea
where $f^{B[M,N]}_{rs}$ and $f^{F[M,N]\pm}_{rs}$ can be found in the Appendix \ref{app twist}.
These rules describe the propagation of individual modes through the twist operator.

\b

(iii) Pair creation: After the modes in the initial state have been either contracted or passed through the twist to become modes after the twist, we are left with the computation of the twist on the initial vacuum state. We have
\bea\label{chi}
|\chi\rangle&=&\sigma^{+}_{2}(w_{0})|0^-\rangle_R^{[M]}|0^-\rangle_R^{[N]}\nn
&=&C_{MN}^{1/2}\,e^{\sum_{s\geq \frac{1}{M+N},r\geq \frac{1}{M+N}}\gamma^{B}_{sr}[\,-\,\alpha_{++,-s}\alpha_{--,-r}\,+\,\alpha_{-+,-s}\alpha_{+-,-r}]}\nn
&&~~~~~~~e^{\sum_{s\geq \frac{1}{M+N},r\geq \frac{1}{M+N}}\gamma^{F}_{sr}[\,d^{++}_{-s}d^{--}_{-r}\,-\,d^{+-}_{-s}d^{-+}_{-r}]}
|0^-\rangle_R^{[M+N]}
\eea
where $C_{MN}$, $\gamma^{B}_{sr}$ and $\gamma^{F}_{sr}$ can be found in the Appendix \ref{app twist}. 
The $\gamma^{B}_{sr}$ describe the creation of  pairs of $\alpha$ modes and the  $\gamma^{F}_{sr}$ describe the creation of  pairs of $d$ modes.

\section{The large $k$ limit}\label{sec large k}

We will be interested in the case where one of the initial component strings has twist $M=k$ and the other has twist $N=1$, to get a final component string with twist $k+1$. Furthermore, we need to understand the amplitude of this process for $k\gg1$. We therefore write the amplitudes found in the above section for this case.

\subsection{The twist operator}\label{sec twist}

We will only keep the $k$ dependence in the limit $k\rightarrow \infty$ with $p,q\sim O(1)$. 

\b

(i) Contraction:

\be\label{c alpha kk}
C\left[\alpha^{[k]}_{A\dot A,-\frac{p}{k}}\,\alpha^{[k]}_{B\dot B,-\frac{q}{k}}\right]\sim \frac{1}{k^2}\epsilon_{AB}\epsilon_{\dot A \dot B}
~~~~~~~~~~p,q>0
\ee
\be\label{c d kk}
C\Big[d^{+ A[k]}_{-\frac{p}{k}}\,d^{-B[k]}_{-\frac{q}{k}}\Big]\sim \frac{1}{k}\epsilon^{AB}
~~~~~~~~~~p\geq 0, q >0
\ee
\be\label{c d k1}
C\Big[d^{-A[k]}_{-\frac{p}{k}}d^{+B[1]}_0\Big]\sim \frac{1}{\sqrt{k}} \epsilon^{AB} 
~~~~~~~~~~ p>0
\ee
where we only consider the modes $d^{+B[1]}_0$ on the singly wound string.

\b

(ii) Propagation: 

\bea
\label{alpha}
\alpha^{[k]}_{A\dot A,-\frac{p}{k}}&\rightarrow& \alpha^{[k+1]}_{A\dot A,-\frac{p}{k+1}}
~~~~~~p>0\\
\label{alpha 1}
\alpha^{[k]}_{A\dot A,-\frac{p}{k}}&\rightarrow& \frac{1}{k}\frac{p}{p-q}\alpha^{[k+1]}_{A\dot A,-\frac{q}{k+1}}
~~~~~p\neq q~~~~p,q>0
\eea
\bea\label{d+}
d^{+A[k]}_{-\frac{p}{k}}&\rightarrow& d^{+A[k+1]}_{-\frac{p}{k+1}}~~~~~~p\geq 0\nn
d^{+A[k]}_{-\frac{p}{k}}&\rightarrow& \frac{1}{k}\frac{q}{p-q} d^{+A[k+1]}_{-\frac{q}{k+1}}
~~~~~p\neq q~~~~p,q\geq 0
\eea
\bea\label{d-}
d^{-A[k]}_{-\frac{p}{k}}&\rightarrow& d^{-A[k+1]}_{-\frac{p}{k+1}}
~~~~~~p>0\nn
d^{-A[k]}_{-\frac{p}{k}}&\rightarrow& \frac{1}{k}\frac{p}{p-q} d^{-A[k+1]}_{-\frac{q}{k+1}}
~~~~~p\neq q~~~~p,q>0
\eea
\bea\label{d+ 1}
d^{+A[1]}_{0}&\rightarrow&\frac{1}{\sqrt{k}} d^{+A[k+1]}_{-\frac{p}{k+1}}~~~~~~p\geq 0
\eea
 Let us call the integer $p$ in the mode $O^{[k]}_{-\frac{p}{k}}$ the `mode number' of the excitation.
From (\ref{alpha}), (\ref{d+}) and (\ref{d-}), we see that the modes in the string with length $k$ tend to keep the same mode number when joining to a string with length $k+1$. Jumping to other mode numbers is suppressed by a relative factor $\sim 1/k$.  From (\ref{d+ 1}), we see that the zero mode on singly wound component string transitions to any mode number on the $k+1$ wound string with a suppression factor $\sim 1/\sqrt{k}$.   

\b

(iii) Pair creation:

\be\label{gamma B}
\gamma^{B}_{\frac{p}{k+1},\frac{q}{k+1}}=-\frac{1}{p+q}
\ee
and
\be\label{gamma F}
\gamma^{F}_{\frac{p}{k+1},\frac{q}{k+1}}=\frac{p}{p+q}\frac{1}{k}
\ee

\b

\subsection{Action of $D$ in the large $k$ limit}\label{sec deformation}

Let us now compute the effect of the deformation operator $D$ on a state containing a single bosonic excitation $\alpha$, in the large $k$ limit with $p,q\sim O(1)$.

For the first term in (\ref{deform left}), we have 
\bea
G^{-}_{\dot A,0}\sigma~\alpha^{[k]}_{B\dot B,-\frac{p}{k}}~\rightarrow~ G^{-}_{\dot A,0}\, f^{B[k]}_{\frac{p}{k},\frac{q}{k+1}} \,\alpha^{[k+1]}_{B\dot B,-\frac{q}{k+1}}
~\rightarrow~f^{B[k]}_{\frac{p}{k},\frac{q}{k+1}}\, \frac{q}{k+1} i \epsilon_{AB}\epsilon_{\dot A\dot B}\,d^{- A[k+1]}_{-\frac{q}{k+1}}
\label{peight}
\eea
In the first step, we have used the propagation rule (\ref{form alpha}). In the second step, we have used the commutator (\ref{app com current a d}). For the second term in (\ref{deform left}), we have 
\bea
\sigma G^{-}_{\dot A,0}~\alpha^{[k]}_{B\dot B,-\frac{p}{k}}~\rightarrow~ \sigma~\frac{p}{k}\, i \epsilon_{AB}\epsilon_{\dot A\dot B}\,d^{- A[k]}_{-\frac{p}{k}}~\rightarrow~f^{F[k]-}_{\frac{p}{k},\frac{q}{k+1}}~ \frac{p}{k} i \epsilon_{AB}\epsilon_{\dot A\dot B}\,d^{- A[k+1]}_{-\frac{q}{k+1}}
\label{pnine}
\eea
In the first step, we used the commutator (\ref{app com current a d}). In the second step, we used the propagation rule (\ref{form d-}).

For the low energy modes $p,q\sim O(1)\ll k$, we have $\frac{p}{k}\neq\frac{q}{k+1}$.
From the (\ref{f M}), we have
\be
f^{F[k]-}_{\frac{p}{k},\frac{q}{k+1}}=f^{B[k]}_{\frac{p}{k},\frac{q}{k+1}}
\ee
Thus we have
\be\label{alpha to d G}
(G^{-}_{\dot A}\sigma)~\alpha^{[k]}_{B\dot B,-\frac{p}{k}}~\rightarrow~-\frac{p}{k^2}i \epsilon_{AB}\epsilon_{\dot A\dot B}\,d^{- A[k+1]}_{-\frac{q}{k+1}}~~~~~~~~p,q>0~~~
\ee

Similarly, we can  compute the effect of the deformation operator $D$ on a state containing a single fermionic excitation $d$, in the large $k$ limit. 
We have
\bea
G^{-}_{\dot A,0}\sigma~d^{+ B[k]}_{-\frac{p}{k}}~\rightarrow~
G^{-}_{\dot A,0}\,f^{F[k]+}_{\frac{p}{k},\frac{q}{k+1}}\,d^{+ B[k+1]}_{-\frac{q}{k+1}}
~\rightarrow~f^{F[k]+}_{\frac{p}{k},\frac{q}{k+1}}\, i \epsilon^{AB}\,\alpha^{[k+1]}_{A\dot A,-\frac{q}{k+1}}\nn
\sigma G^{-}_{\dot A,0}~d^{+ B[k]}_{-\frac{p}{k}}~\rightarrow~\sigma ~i \epsilon^{AB}\alpha^{[k]}_{A\dot
A,-\frac{p}{k}}~\rightarrow~ f^{B[k]}_{\frac{p}{k},\frac{q}{k+1}} \,i \epsilon^{AB}\,\alpha^{[k+1]}_{A\dot A,-\frac{q}{k+1}}
\label{ptenq}
\eea
From (\ref{f B}) and (\ref{f M}), at  leading order in the large $k$ limit, we have
\be
f^{F[k]+}_{\frac{p}{k},\frac{q}{k+1}}-f^{B[k]}_{\frac{p}{k},\frac{q}{k+1}} = -\frac{1}{k}  ~~~~~~~~p\geq 0~~q>0
\ee
Thus we have
\be\label{G d+ k}
(G^{-}_{\dot A}\sigma)~d^{+ B[k]}_{-\frac{p}{k}}~\rightarrow~-\frac{1}{k} i \epsilon^{AB}\alpha^{[k+1]}_{A\dot A,-\frac{q}{k+1}}
~~~~~~~~p\geq 0~~q>0
\ee

Finally we consider the  fermionic zero mode $d^{+B(1)}_{0}$ in the singly wound string, in the large $k$ limit. We have
\bea\label{G d+ 1}
(G^{-}_{\dot A}\sigma)~d^{+B[1]}_{0}&\rightarrow&\frac{1}{\sqrt{k}} i \epsilon^{AB}\alpha^{[k+1]}_{A\dot A,-\frac{q}{k+1}}
\eea

\subsection{A heuristic derivation of the large $k$ limit}\label{sec heuristic k}

In the above computations we have derived the large $k$ behavior of the action of the deformation operator $D$ by starting with the exact expressions for the action of $D$. Given the importance of the powers of $k$ arising in the resulting expressions, we give here a heuristic understanding of this $k$ dependence. The $k$ dependence can be understood from the following set of rules:

\b

(a) Consider a fermionic mode $d_{-{n_1\over k}}$ on the $k$ wound string. We assume that $n_1\sim O(1)$, while $k\gg 1$. The fermionic mode defined in (\ref{i string modes}) has some part of its support on each of the $k$ strands of the string. Thus we can schematically write the excitation as
\be
d_{-{n_1\over k}}\r {1\over \sqrt{k}} \left ( e^{2\pi i {n_1\over k}}d^{(1)} + e^{4\pi i {n_1\over k}}d^{(2)} +\dots e^{2\pi i k{n_1\over k}}d^{(k)} \right ) 
\label{kone}
\ee
where $d^{(i)}$ is a fermionic excitation localized near string $i$. A second mode $d_{-{n_2\over k}}$ has a similar decomposition. 

Now consider the contraction between these two modes under the action of the twist operator $\sigma_2$.   Suppose the twist operator $\sigma_2$ joins strand $1$ of the $k$-wound string  with a singly wound string. The effect of the twist is localized to a region of length $\sim 2\pi$; i.e. the changes to the state occur (with all parameters of order unity) in the region extending to around $\sim 1$ windings around the point of insertion of the twist.  Thus only the parts $d^{(1)}$ in expressions like (\ref{kone}) contribute to the contraction. The prefactor ${1\over \sqrt{k}}$ in (\ref{kone}) from the two modes then gives a factor ${1\over \sqrt{k}}{1\over \sqrt{k}}\sim {1\over k}$, which agrees with the $k$ dependence seen in (\ref{c d kk}).

In the contraction (\ref{c d k1}) we have one $d$ mode on the $k$ wound string, which supplies a factor ${1\over \sqrt{k}}$, and another $d$ mode from the singly wound string, which has no such factor. Thus we recover the behavior $\sim {1\over \sqrt{k}}$ in (\ref{c d k1}).

\b

(b) The bosonic modes behave in a similar way, except that the $\alpha$ operators are normalized differently from the usual creation operators:
\be
\alpha_{-{n\over k}} \r \sqrt{n\over k} a^\dagger_{n\over k}
\label{pfive}
\ee
With $n\sim O(1)$ and $k\gg1$, we see that the contraction between two $\alpha$ modes has an extra factor ${1\over \sqrt{k}}{1\over \sqrt{k}}\sim {1\over k}$ compared to the contraction of fermion modes, and so we recover the $\sim {1\over k^2}$ behavior seen in (\ref{c alpha kk}). 

\b

(c) Now consider the propagation of a mode $d_{-{n_1\over k}}$ on the $k$ wound string to a mode on the $k+1$ wound string. Before the twist, this mode can be schematically represented as (\ref{kone}). After the twist, the singly wound string is added on to the $k$ wound string. The state of the fermion is altered in the loop corresponding to this singly wound string and in the loop (which we had taken to be loop $1$) on the $k$ wound string where the twist acted. Thus the state on the $k+1$ wound string has the form
\be
|\psi\rangle={1\over \sqrt{k}} \left ( \beta_0 d^{(0)}  + \beta_1 d^{(1)} + e^{4\pi i{n_1\over k} }d^{(2)} +\dots e^{2\pi ik{n_1\over k}} d^{(k)} \right ) 
\label{pthree}
\ee
where $d^{(0)}$ is the fermion state on the loop coming from the singly wound string, and $d^{(i)}, i=1, \dots k$ are  fermions on loop $i$ of the $k$ wound string as before. The states of the fermion on loops $2, 3, \dots k$ are unaltered by the twist in our approximation. Note that the parameters $\beta_0$ and  $\beta_1$ are of order unity.

Now we take the dot product of this state with the Fourier mode $d_{-{n_1\over k+1}}$ on the $k+1$ wound string. This mode has the form (with a convenient choice of overall phase)
\be
|\psi_{-{n_1\over k+1}}\rangle = {1\over \sqrt{k+1}}\left ( d^{(0)}  +  e^{2\pi i{n_1\over k+1} } d^{(1)} + e^{4\pi i{n_1\over k+1} } d^{(2)} +\dots  e^{2\pi ik{n_1\over k+1} }d^{(k)} \right ) 
\label{pfour}
\ee
Since ${n_1\over k}-{n_1\over k+1}\sim {n_1\over k^2}$ is very small, we see that the overlap between (\ref{pthree}) and (\ref{pfour}) is order unity. This gives the first relations in (\ref{d+}) and (\ref{d-}). On the other hand if we let the mode on the $k+1$ wound string be $d_{-{n_2\over k+1}}$ with $n_2\ne n_1$, then then (\ref{pfour}) is close to orthogonal to (\ref{pthree}); the nonzero part of the overlap comes from the first two terms in (\ref{pthree}) which differ by order unity from the terms in the Fourier mode $d_{-{n_1\over k}}$. These terms give a contribution $\sim {1\over \sqrt{k}}{1\over \sqrt{k}}\sim {1\over k}$ from the prefactors in (\ref{pthree}) and (\ref{pfour}), giving the $\sim {1\over k}$ behavior in the second relations in (\ref{d+}) and (\ref{d-}).

The bosonic relations (\ref{alpha}) are reproduced in a similar way; the normalization (\ref{pfive}) does not enter these relations as it cancels out on the two sides of the propagation relation.

The relation (\ref{d+ 1}) is obtained by noting that the Fourier mode in the final state has the prefactor $\sim {1\over \sqrt{k}}$ from (\ref{pfour}) but there is no such prefactor from the normalization of the mode on the singly wound string. Thus we get just the dependence $\sim {1\over \sqrt{k}}$. 

\b

(d) The pair creation relation (\ref{gamma F}) for fermions can be understood in a similar way. Let the twist act in the first loop of the $k$ wound string. Then the pair of fermions is created in a region of width $\sim 2\pi$ around the twist insertion. Taking the inner product of each fermion with a Fourier mode of the form (\ref{pfour}) gives a factor $\sim {1\over \sqrt{k}}{1\over \sqrt{k}}\sim {1\over k}$, as seen in (\ref{gamma F}). 

The bosonic relation (\ref{gamma B}) is obtained the same way, after accounting for the normalization change (\ref{pfive}):
\be\label{pair boson}
{1\over k}  a^\dagger_{{n_1\over k}} a^\dagger_{{n_2\over k}}\sim {1\over k} \sqrt{k\over n_1}\sqrt{k\over n_2}\alpha_{-{n_1\over k}}\alpha_{-{n_2\over k}}\sim  \alpha_{-{n_1\over k}}\alpha_{-{n_2\over k}}
\ee
which gives the $k$ dependence of (\ref{gamma B}), assuming that $n_1, n_2\sim O(1)$.

\b

(e) In supersymmetry we have a supercharge $G$ which squared to the energy $G^2\sim H$. For our left movers, $H$ is the left momentum. For our case where we have  modes ${n\over k}$ with $n\sim O(1), k\gg 1$, this momentum is $\sim {1\over k}$. Thus a supercharge converts a bosonic creation operator $a^\dagger_{n\over k}$ to a fermionic creation operator $d_{-{n\over k}}$, or vice versa, with a factor $\sim {1\over \sqrt{k}}$.  Writing this in  terms of $\alpha$ oscillators using (\ref{pfive}), we find that
\be\label{alpha to d}
G_0\,  \alpha_{-{n\over k}}\sim{1\over \sqrt{k}} G_0\,  a^\dagger_{{n\over k}} \sim {1\over \sqrt{k}}{1\over \sqrt{k}} d_{-{n\over k}}\sim {1\over k} d_{-{n\over k}}
\ee
\be\label{d to alpha}
G_0\,  d_{-{n\over k}}\sim {1\over \sqrt{k}}\, a^\dagger_{{n\over k}}  \sim {1\over \sqrt{k}}{\sqrt{k}}\, \alpha_{-{n\over k}}\sim \alpha_{-{n\over k}}
\ee
the first of these relations explains the $k$ powers in (\ref{peight}), (\ref{pnine}), and the second explains the powers of $k$ in the relations (\ref{ptenq}). 

We also see that the application of $G_0$ before and after a twist gives almost the same state, so that the difference in contours (\ref{deform left}) brings in a further factor ${1\over k}$. This can be understood by writing the supercharge on a state like (\ref{kone}) in an approximate form as
\be
G_0\r G_0^{(1)}+G_0^{(2)}\dots G_0^{(k)}
\ee
where $G_0^{(i)}$ acts on copy $i$ to convert a boson on that copy to a fermion or vice versa. We see  that the difference between $G_0$ acting before or after the twist arises only from the components $d^{(0)}, d^{(1)}$, and the prefactors $\sim {1\over \sqrt{k}}$ in the initial and final state generate the suppression ${1\over k}$ when we take the difference between supercharge modes in  (\ref{deform left}).

In the relation (\ref{G d+ 1}), the action of $G_0$ before the twist annihilates the $d_0$ mode, so we have only the action of $G_0$ after the twist. 

\section{Evolution of excitations on CFT states $\left ( |0\rangle^{[1]}_R |\bar 0\rangle^{[1]}_R \right )^{N}$}\label{sec T4 special}

With the above computations and estimates, we now address our issue. In this section we discuss the evolution of a perturbation on the CFT state of type (\ref{cft2}). In the gravity description such a state originates from an NS1P profile like (\ref{vib2}) which has vibrations along the compact $T^4$;  such geometries have deep throats and thus very small energy gaps. Our issue is to reproduce this small energy gap in the CFT. At the orbifold point the energy gap is (\ref{ecft2}) {\it not} small, so we have to look at the effect of the deformation operator $D$ and see how it converts the large gap to a small gap.

To make the issue most concrete, let us take the CFT state at the orbifold point to be in the untwisted sector. We must now choose the fermion zero modes to get the state of each copy of the $c=6$ CFT. For the case of interest where the starting NS1P  had vibration profiles along the $T^4$, the left part of CFT state can be in one of the two vacua $|0\rangle_R^{[1]}$, $|\t 0\rangle_R^{[1]}$. Similarly, the right part of the CFT state can be in one of the two vacua $| \bar 0\rangle_R^{[1]}$, $|\bar {\t 0}\rangle_R^{[1]}$. We choose all our copies to be in the state
\be
|\psi\rangle^{[1]}= |0\rangle_R^{[1]}|\bar 0\rangle_R^{[1]} 
\ee
where the superscript $[1]$ tells us that the copy of the $c=6$ CFT has twist $1$, i.e., it is in the untwisted sector. We can write the above state over a Ramomd ground state with lower charges, by explicitly writing out the fermion zero modes that cancel the charges: 
\be
|\psi\rangle^{[1]}=d^{++[1]}_0\bar d^{++[1]}_0 |0^-\rangle_R^{[1]}|\bar 0^-\rangle_R^{[1]}
\ee
where again the superscripts $[1]$ on the fermion operators tell us that they are acting on a CFT copy with twist $1$. 

Since we have $N=n_1n_5$ copies of the $c=6$ CFT, the overall state of the CFT has the form
\be
|\Psi\rangle=\Big (d^{++[1]}_0\bar d^{++[1]}_0 |0^-\rangle_R^{[1]}|\bar 0^-\rangle_R^{[1]}\Big )^{N}
\label{unpert}
\ee
where we have $N$ factors describing the $N$ copies of the $c=6$ CFT.

We take a simple excitation of this state, where we place  a pair of excitations on the first copy of the $c=6$ CFT
\bea
|\psi\rangle=\Big (\alpha_{-+, -1}^{[1]}\bar\alpha_{-+, -1}^{[1]}d^{++[1]}_0\bar d^{++[1]}_0 |0^-\rangle_R^{[1]}|\bar 0^-\rangle_R^{[1]}\Big )\Big (d^{++[1]}_0\bar d^{++[1]}_0 |0^-\rangle_R^{[1]}|\bar 0^-\rangle_R^{[1]}\Big )^{N-1} \label{initial}
\eea
The left and right excitation energies are
\be
 h=1, ~~~ \bar h=1
\label{energies}
\ee
In this description at the orbifold point, we see that the energy gap for excitations (with zero total momentum) on the state (\ref{unpert})  is
\be
\Delta E_{CFT}=2
\label{gap1}
\ee
which becomes the expression (\ref{ecft2}) for $k=1$ after we rescale the $S^1$ from having length $2\pi$ to its physical length $2\pi R_y$. Our goal is to examine the effect of the deformation operator $D$ on  (\ref{initial}); we will find that the large energy gap (\ref{gap1}) gets converted to a small energy gap by the emergence of a band structure.

In section \ref{sec join} we outline the steps in the computation, and in section \ref{sec amplitude} we carry out the estimates needed in these steps.

\subsection{Eliminating the fermionic modes by supercharges}\label{sec convert}

The twist operator in the deformation operator $D$ acting on the state (\ref{initial}) generates a state that is an infinite linear combination of different energy eigenstates of the orbifold theory. 
We have started with string $1$ having the excitation
\be
\alpha_{-+, -1}^{[1]}\bar\alpha_{-+, -1}^{[1]}d^{++[1]}_0\bar d^{++[1]}_0 |0^-\rangle_R^{[1]}|\bar 0^-\rangle_R^{[1]}
\ee
Suppose the twist $\sigma_2$ in the deformation operator joins this string to string 2, which is in the ground state
\be
d^{++[1]}_0\bar d^{++[1]}_0 |0^-\rangle_R^{[1]}|\bar 0^-\rangle_R^{[1]}
\ee
Let us ask: what happens to the fermionic modes $d^{++}_0, \bar d^{++}_0$ after the deformation operator joins strings $1$ and $2$ into a string with winding $k=2$? We noted in section \ref{sec twist} that under the action of the twist $\sigma_2$ there were two possibilities for a fermion mode in the initial state: contraction with another fermionic mode, propagation through the twist. The fermionic zero modes we are discussing have positive $SU(2)_L, SU(2)_R$ charges, and we see that there are no fermions with negative charges that they can contract with. The modes can propagate through the twist to give fermionic modes on the $k=2$ twisted string. In later steps we will join this string with more singly wound strings, to generate strings with winding $k\gg 1$. Each singly wound string of type $d^{++[1]}_0\bar d^{++[1]}_0 |0^-\rangle_R^{[1]}|\bar 0^-\rangle_R^{[1]}$ brings in an extra pair of fermions $d^{++}_0, \bar d^{++}_0$. Suppose we collect $m$ such fermions on the final $k$ twisted string. By the Pauli exclusion principle, we cannot put two of these fermions in the same mode on this string. The energy levels on the $k$ twisted string are in multiples of $1/k$, so the minimum energy we need to have $m$ fermions $d^{++}$ on it is
\be\label{sea energy}
h= {0\over k}+{1\over k}+{2\over k} +\dots {m-1\over k}={m(m-1)\over 2 k} \sim {m^2\over k}
\ee
But the total excitation energy in the left sector that we have started with in (\ref{initial}) is $ h=1$. 
Assuming energy conservation in the evolution process, we see that the number $m$ of fermionic modes on the $k$ wound string must satisfy 
\be
m\lesssim \sqrt{k} \ll k
\ee
Thus most of the fermionic zero modes $d^{++}_0, \bar d^{++}_0$ from $k$ singly wound strings have to be removed from the final $k$ wound string. Let us assume that the fermion  mode is indeed removed in some way. How do we accomplish this?

The deformation operator $D$ has supercharges $G_0, \bar G_0$ acting on the left and right sectors; each of these supercharges can either act before the twist or after the twist. A $G_0$ before the twist annihilates a mode $d_0$, so we get no contribution from this. A $G_0$ mode after the twist can map a mode $d_{-{n\over k}}$ to a mode $\alpha_{-{n\over k}}$. The overall effect is (\ref{G d+ 1})
\bea\label{G d+ 1 s}
(G^{-}_{\dot A}\sigma)~d^{++[1]}_{0}&\rightarrow&\frac{1}{\sqrt{k}} i \alpha^{[k+1]}_{-\dot A,-\frac{q}{k+1}}
\eea
Bosonic modes $\alpha$ do not satisfy a Pauli exclusion principle, so we can have an arbitrarily large number of them in a low energy modes on the $k$ wound string. This will give us the process: each time we join a singly wound string to our $k$ wound string, we get an extra fermionic mode $d_0^{++}$; this propagates through the twist to a mode $d^{++}_{-{n\over k+1}}$, and the $G_0$ contour after the deformation converts this to a bosonic mode. In this way, the fermionic zero modes from singly wound strings can be eliminated by supercharges.

\subsection{The process of joining short strings into a long string}\label{sec join}

Let us now determine the mode number $q$ of the bosonic mode in (\ref{G d+ 1 s}). When we go from the initial state (\ref{initial}) to the state with $k=2$, we need one extra boson on each of the left and right sides. Thus we now have a total of two bosons on each of the left and right sides. We can maintain energy conservation by placing each of these bosons in the lowest allowed mode $1/2$. Similarly, when we have joined $k-1$ singly wound strings to our initial string, we have a total winding $k$, and $k$ bosons on each of the left and right sides. Energy conservation can be maintained by placing each of these bosons in the lowest allowed mode $1/k$. Now consider the process to get a component string of winding $k+1$. As shown in (\ref{alpha}), these $k$ bosons in the mode $1/k$ can propagate through the twist and become $k$ bosons in the mode $1/(k+1)$ without suppression. Thus to maintain energy conservation, the mode number $q$ of the new created bosonic modes in (\ref{G d+ 1 s}) must be $q=1$.

Note that all the  bosons $\alpha_{A \dot A}$ on the multiwound string have $A$ charge equal to $-1/2$ as indicated in (\ref{G d+ 1 s}). Thus there cannot be any contraction between these bosons at any step of the evolution. Further, we cannot generate other bosons by the process of pair creation while still maintaining energy conservation; this follows because we have already exhausted our energy budget of $1$ unit on each of the left and right sides by placing the required number of bosonic modes in their lowest allowed energy states.   

Let us now write the states resulting from the above process in more detail, keeping track of spins and numerical factors. Suppose we are at a state where we have twist $k$. There are $k$ bosonic operators on this twisted string, each in the mode $1/ k$. These bosonic modes come from the fermionic modes of the singly wound strings through (\ref{G d+ 1 s}), so they can be $\alpha^{[k+1]}_{-\pm,-\frac{1}{k+1}}$. Consider a state of a $k$-wound string
\be\label{T4 k string}
|k; q_L, q_R\rangle\equiv (\alpha^{[k]}_{-+,-{1\over k}})^{q_L}(\alpha^{[k]}_{--,-{1\over k}})^{k-q_L}
(\bar \alpha^{[k]}_{-+,-{1\over k}})^{q_R}(\bar \alpha^{[k]}_{--,-{1\over k}})^{k-q_R}|0\rangle_R^{[k]}|\bar 0\rangle_R^{[k]} 
\ee
For generic states we have
\be
q_L, ~k-q_L, ~q_R,~ k-q_R~~\sim ~~k
\ee
The deformation operator $D$ twists this string with a singly wound string to yield a $(k+1)$-wound string. 
The left mover of the singly wound string $|0\rangle_R^{[1]}|\bar 0\rangle_R^{[1]} $ is
\be
|0\rangle_R^{[1]}=d^{++[1]}_{0}|0^-\rangle_R^{[1]}
\ee
In the process, the fermionic mode $d^{++[1]}_{0}$ is converted to a bosonic mode. 
Using (\ref{G d+ 1}), we get
\bea\label{T4 alpha}
(G^{-}_{ +}\sigma)~d^{++[1]}_{0}&\rightarrow&\frac{1}{\sqrt{k}} i \alpha^{[k+1]}_{-+,-\frac{1}{k+1}}~=~ \frac{1}{\sqrt{k(k+1)}} i a^{\dagger[k+1]}_{-+,\frac{1}{k+1}}\nn
(G^{-}_{ -}\sigma)~d^{++[1]}_{0}&\rightarrow&\frac{1}{\sqrt{k}} i \alpha^{[k+1]}_{--,-\frac{1}{k+1}}~=~ \frac{1}{\sqrt{k(k+1)}} i a^{\dagger[k+1]}_{--,\frac{1}{k+1}}
\eea
where we have written the $\alpha$ modes in terms of creation operators $a^\dagger$ using the relation (\ref{pfive}). Similar process happens for the right mover.
The deformation operator (\ref{D 1/4}) has the form
\be
\label{jjfour}
D= - D_+ \bar D_-+D_- \bar D_+
\ee
where the subscripts give the $\dot A$ index on the respective operators.
Thus we obtain one of the states $|k+1; q_L+1, q_R\rangle$ or $|k+1; q_L, q_R+1\rangle$. 

\subsection{The amplitude in the large $k$ limit}\label{sec amplitude}

In this section, we will compute the amplitude
\be\label{T4 amp s}
|k; q_L, q_R\rangle |0\rangle_R^{[1]}|\bar 0\rangle_R^{[1]}  \rightarrow |k+1; q_L+1, q_R\rangle ~~\text{or}~~ |k+1; q_L, q_R+1\rangle
\ee
in the large $k$ limit, where states are defined by
\be\label{T4 k string 1}
|k; q_L, q_R\rangle\equiv (\alpha^{[k]}_{-+,-{1\over k}})^{q_L}(\alpha^{[k]}_{--,-{1\over k}})^{k-q_L}
(\bar \alpha^{[k]}_{-+,-{1\over k}})^{q_R}(\bar \alpha^{[k]}_{--,-{1\over k}})^{k-q_R}|0\rangle_R^{[k]}|\bar 0\rangle_R^{[k]} 
\ee
and
\be
|0\rangle_R^{[k]}|\bar 0\rangle_R^{[k]} 
=d^{++[k]}_0\bar d^{++[k]}_0|0^-\rangle_R^{[k]}|\bar 0^-\rangle_R^{[k]}
\ee
We will first look at the left movers and then combine the left and right movers together.
We proceed in the following steps.

\b

(a) Twisting the Ramond ground states: 
The twist operator can twist the Ramond ground states $|0^-\rangle_R^{[k]}$ and $|0^-\rangle_R^{[1]}$ into a Ramond ground state of the $(k+1)$-wound string. In the large $k$ limit, this gives an order $O(1)$ factor from (\ref{chi})
\be\label{T4 twist ground}
\sigma^{+} |0^-\rangle_R^{[k]}  |0^-\rangle_R^{[1]} \rightarrow  \frac{1}{\sqrt 2} |0^-\rangle_R^{[k+1]}  
\ee
In the state $|k; q_L, q_R\rangle$ (\ref{T4 k string 1}), there is an extra fermionic zero mode in $|0\rangle_R^{[k]}=d^{++[k]}_0 |0^-\rangle_R^{[k]}$ compared to the $|0^-\rangle_R^{[k]}$ in eq.\,(\ref{T4 twist ground}).
This fermionic zero mode propagates through the twist operator and become a fermionic zero modes in the $(k+1)$-wound string from the first equation in eq.\,(\ref{d+}). Thus we have
\be\label{T4 twist ground 2}
\sigma^{+} |0\rangle_R^{[k]}  |0^-\rangle_R^{[1]} \rightarrow  \frac{1}{\sqrt{2}} |0\rangle_R^{[k+1]}  
\ee
The fermionic zero mode in the singly wound string will be considered in the step (c).

\b

(b) Propagation of bosonic modes: The $k$ bosons in the mode $1/k$ in the $k$-wound string become $k$ bosons in the mode $1/(k+1)$ in the $(k+1)$-wound string.
\be\label{bosons T4}
\sigma (\alpha^{[k]}_{-+,-{1\over k}})^{q_L}(\alpha^{[k]}_{--,-{1\over k}})^{k-q_L} \rightarrow ~O(1)~(\alpha^{[k+1]}_{-+,-{1\over k+1}})^{q_L}(\alpha^{[k+1]}_{--,-{1\over k+1}})^{k-q_L}
\ee
From eq.\,(\ref{alpha}), we find the amplitude for each boson is $1+O(k^{-1})$. Thus the amplitude for $k$ bosons is at order $O(1)$. From eq.\,(\ref{alpha 1}), we find that changing the  `mode number' of these bosons is suppressed by extra factors of $1/k$. 

\b

(c) Eliminating the fermionic modes by supercharges: 
As explained in eq.\,(\ref{T4 alpha}), the amplitude to convert the fermionic zero mode $d^{++[1]}_{0}$  in the singly wound string to a bosonic mode $\alpha^{[k+1]}_{-\pm,-\frac{1}{k+1}}$  in the $(k+1)$-wound string is at order
\be\label{T4 convert}
\frac{1}{k}
\ee

\b

(d) The Bose enhancement factor: If there are $k$ bosonic excitations  already in a normalized state, then application of a creation operator in the same mode would yield an extra factor
$
\sqrt{k+1}
$, which can be seen from
\be\label{add boson}
a^\dagger \left[ \frac{1}{\sqrt{k!}}(a^\dagger)^k |\Omega\rangle \right]=\sqrt{k+1} \left[\frac{1}{\sqrt{(k+1)!}}(a^\dagger)^{k+1}|\Omega \rangle\right]
\ee
where the states in the square brackets are normalized. The creation and annihilation operators are defined through $[a,a^{\dagger}]=1$ and $a|\Omega\rangle=0$. Similarly, removal of a creation operator by applying an annihilation would yield a factor $\sqrt{k}$
\be\label{remove boson}
a \left[\frac{1}{\sqrt{k!}}(a^\dagger)^k |\Omega\rangle\right]=k \frac{1}{\sqrt{k!}}(a^\dagger)^{k-1}|\Omega \rangle
=\sqrt{k} \left[ \frac{1}{\sqrt{(k-1)!}}(a^\dagger)^{k-1}|\Omega \rangle\right]
\ee
These are the so-called Bose enhancement factors.

The step (b) add a bosonic mode $\alpha^{[k+1]}_{- -,-\frac{1}{k+1}}$ or $\alpha^{[k+1]}_{- +,-\frac{1}{k+1}}$ to the state (\ref{bosons T4}). The number of these bosonic modes in (\ref{bosons T4})  are of order $\sim k$. Thus the step (b) also contribute a Bose enhancement factor 
\be\label{T4 bose}
\sqrt{k}
\ee

\b

(e) Combining the left and right movers:  From eqs.\,(\ref{T4 twist ground 2}), (\ref{bosons T4}), (\ref{T4 convert}) and (\ref{T4 bose}), we have the amplitude for the left mover is
\be
\label{jjfive}
A_L\sim A_R \sim \frac{1}{\sqrt{k}}
\ee
Similar result holds for the amplitude $A_R$ of right movers.
We include an extra factor $k$ when combining the left and right amplitudes; this extra factor arises from the fact that the singly wound string can twist with any of the $k$ copies in the $k$-twisted string. Then we find that the full amplitude is
\be\label{T4 amp}
A = \lambda~k~A_L~A_R \sim \lambda
\ee
where $\lambda$ is the coupling defined in (\ref{defor S}). Note that this amplitude for transition from the $k$ wound component string to the $k+1$ wound component string does not decrease as we go to larger $k$. 

\section{Band structure}\label{sec band}

Let us now come to the spectrum of excitations. At the orbifold point, the excitations around the singly wound sector state (\ref{unpert}) are spaced with
\be
\Delta E = {2}
\label{jjone}
\ee
where we have scaled the CFT circle to have length $2\pi$. Such an anergy gap does not describe a deep throat of the kind seen in the dual geometry. We will now see how the action of the deformation operator $D$ converts this widely spaced spectrum into a band of closely spaced states; such a band can indeed describe the closely spaced spectrum seen in the dual gravity theory.

We have seen that the deformation operator leads to a transition from the initial state (\ref{initial}) to the states in the set (\ref{T4 k string}). We start at $k=1$, and at each step in the evolution we increase $k$ by $1$. This is reminiscent of an electron hopping along a set of atoms forming a crystal lattice. The energy of the states (\ref{T4 k string}) is the same at each value of $k$, so  all atoms in the lattice are at the same energy. The transition from $k$ to $k+1$ is like the `hopping' term which allows the electron to make a transition from one lattice site to a neighboring site. 

Because of the spin indices in the states (\ref{T4 k string}), we will see that the lattice formed by these states is a 2-dimensional one. But to understand the band formation, let us first recall how a band emerges from the existence of a hopping interaction in a 1-dimensional lattice.

\subsection{Band structure in 1-dimension}

Consider a 1-dimensional lattice of $N$ points, with the spacing between points being $1$. For convenience we let the last point of the lattice be linked to the first, to make a periodic system; our actual lattice will not be periodic, but the qualitative picture of band formation is the same with periodic or vanishing boundary conditions. 

A particle can sit at any of these sites with the same energy; without loss of generality we can set this energy to zero.  A hopping term $\epsilon$  gives the amplitude per unit time for transition from lattice site $k$ to the neighboring sites $k+1$ and $k-1$. The nonzero terms in the Hamiltonian $\hat H$ are given by
\be
H_{k, k+1}=H_{k+1,k}=\epsilon
\ee
with $k=N+1$ being identified with $k=1$. 

The eigenvectors of this Hamiltonian are
\be
\psi_s(k) ={1\over \sqrt{N}} e^{isk}, ~~~s={2\pi m\over N}, ~~m=0, 1, \dots N-1
\label{jjthree}
\ee
with eigenvalues
\be
E_s=2 \epsilon \cos s
\ee
Note that 
\be
0\le s< 2\pi
\ee
and there are $N$ eigenvalues in this range. Thus for $N$ large, the spacing of eigenvalues is very small, and we can describe them by a band. In particular, the spacing is much less than (\ref{jjone}). This band formation is the effect of the hopping term. In our problem the hopping arises from the deformation operator $D$ that takes us off the orbifold point towards the supergravity point.

\subsection{Band structure in 1-dimension with variable hopping strength}\label{secnonconstant}

Before we come to our actual D1D5 system, let us study an extension of the 1-dimensional lattice considered above. Suppose the hopping term $\epsilon$ was not the same everywhere on the lattice, but instead varied slowly with position along the lattice. Between the sites $k, k+1$ we write the hopping term as $\epsilon_{k, k+1}$. Here `slowly varying' means that between one neighboring pair of sites and the next,
\be
{\delta \epsilon\over \epsilon} \ll 1
\ee
Consider an eigenfunction of the Hamiltonian with energy $E$. 
\be
\psi_E=\{ \psi_1, \psi_2, \dots \psi_k, \dots \psi_N\}
\ee
Consider the following relation from the eigenvalue equation $\hat H \psi_E=E\psi_E$
\be
\sum_j (\hat H)_{kj}\psi_j=E\psi_k
\ee
Since the hopping terms are the only nonzero terms in $\hat H$, this relation is
\be
\epsilon_{k-1,k} \psi_{k-1}+\epsilon_{k,k+1}\psi_{k+1}=E\psi_k
\label{jjtwo}
\ee
Since the hopping tern is slowly varying, we write 
\be
\epsilon_{k-1,k}\approx \epsilon_{k,k+1}\approx \epsilon(k)
\ee
Adding a shift to both sides of (\ref{jjtwo}) we get
\be
\epsilon(k) \left ( \psi_{k-1}-2\psi_k +\psi_{k+1}\right ) = \left (E-2\epsilon(k)\right ) \psi_k
\ee
For a wavefunction where $\psi_k$ varies slowly with $k$, we can see the LHS of the above equation as a second derivative acting on a function $\psi(k)$. This gives
\be
\epsilon(k) {d^2 \psi(k)\over dk^2} = \left (E-2\epsilon(k)\right ) \psi(k)
\ee
This can be written in the form
\be
-\h  {d^2 \psi(k)\over dk^2} +{E\over 2\epsilon(k)}\psi(k)= \psi(k)
\ee
This is the quantum mechanical eigenvalue  equation for a nonrelativistic particle of unit mass in the potential ${E\over 2\epsilon(k)}$, with eigenvalue $1$. For the case where $\epsilon$ is constant, we get the equation for a free particle, which has the eigenfunctions (\ref{jjthree}) we found above. For slowly varying $\epsilon$, we have the equation of a particle in a slowly varying potential, where we again expect band formation. 

\subsection{Band formation for the D1D5 CFT}

Let us now come to our actual problem.  The deformation operator $D$ creates the twist, which is responsible for the `hopping' from one state to the next. This operator has the form (\ref{D 1/4}). While $D$ is an overall singlet of all the charges in the theory, the $\dot A$ charge is nonzero for each of the left and right sectors; it is the overall operator that is neutral. Thus $D$ can be written in the form (\ref{jjfour}), where the term $D_+\bar D_-$ increases the $\dot A$ charge by $\h$ for the left movers and decreases it by $\h$ for the right movers, and the term $D_-\bar D_+$ decreases the $\dot A$ charge by $\h$ for the left movers and increases it by $\h$  for the right movers. 

\begin{figure}[htbp]
\begin{center}
\includegraphics[scale=0.5]{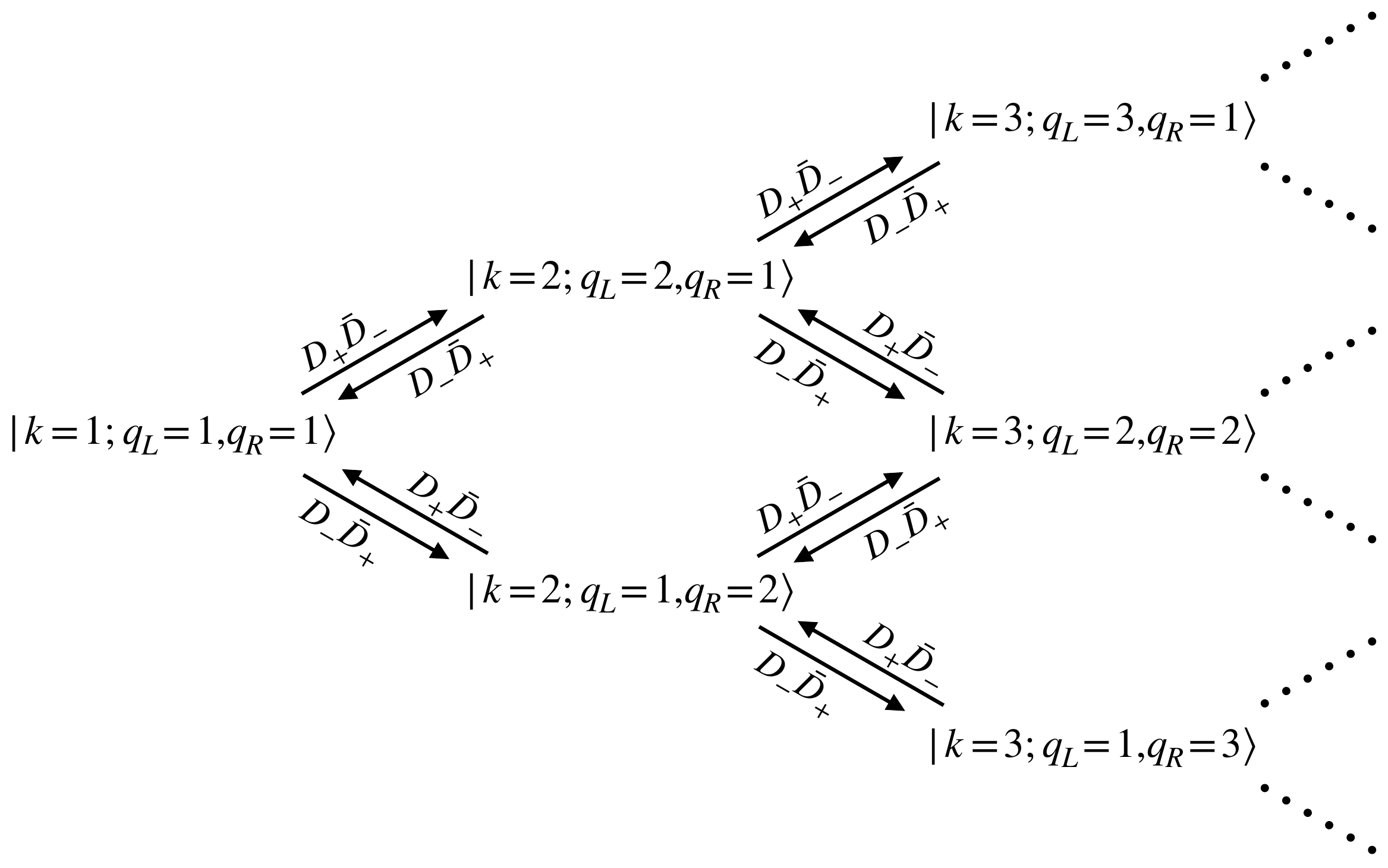}\hskip20pt
\end{center}
\caption{A 2-dimensional lattice with a hopping interaction.}
\label{fig3}
\end{figure}

Thus consider any of the states (\ref{T4 k string}). The index $k$ labels the twist sector, and the deformation operator $D$ changes $k$ by one unit. But it can move us in two directions: (i) increase the $\dot A$ charge for the left mover while decreasing it for the right mover, and (ii) decrease the $\dot A$ charge for the left mover while increasing it for the right mover. Thus the hopping takes place on a two-dimensional lattice as depicted in fig.\ref{fig3}. 

The energy of each of the states (\ref{T4 k string}) is the same; thus we can map to a quantum mechanical model where a particle has the same energy at all lattice sites. We have also seen that the amplitude for hopping (\ref{T4 amp}) does not scale with $k$ for large $k$. But this hopping amplitude is not constant; it varies slowly over the space of states (\ref{T4 k string}). Recall that there is a Bose enhancement factor (\ref{T4 bose}) when we create a new $\alpha$ excitation when moving from twist $k$ to twist $k+1$. Consider the left movers. The $\alpha$ excitations can be of two types: $\alpha_{-+}$ and $\alpha_{--}$. Suppose at twist $k$ we have, for the left movers,  $q_L$
modes $\alpha_{-+, -{1\over k}}$ and $k-q_L$ modes $\alpha_{--, -{1\over k}}$. The Bose enhancement factor for moving in the direction $q_L\r q_L+1$ (by creation of a $\alpha_{-+, -{1\over k+1}}$ excitation) is
\be
\sqrt{q_L+1}
\ee
The Bose enhancement factor for moving in the direction $q_L\r q_L$ (by creation of a $\alpha_{--, -{1\over k+1}}$ excitation) is
\be
\sqrt{k-q_L+1}
\ee
For large $k$, we have $q_L\sim k$. In our initial estimate (\ref{jjfive})  we had set the number of $\alpha_{-+}$ and $\alpha_{--}$ oscillators to be approximately equal. But we see that as $q_L$ varies over the space of states  $(\ref{T4 k string})$, there is in fact a slow variation of the Bose enhancement factors in the amplitude for hopping. 

Thus we get the situation of section \ref{secnonconstant} where we can map the problem to the eigenvalue equation for a nonrelativistic particle in a slowly varying potential, but now in 2-dimensions instead of one. 
In this situation we again expect the formation of a band, and thus we get the closely spaced levels required to match the fuzzball states in the dual gravity description: these fuzzball states have a deep throat, and thus a closely spaced spectrum of excitations.

\section{States dual to short throat geometries $\left(|0^+\rangle_R^{[1]}|\bar 0^+\rangle_R^{[1]}\right)^N$}\label{sec X4}

In sections \ref{sec T4 special} and \ref{sec band}, we have studied the CFT states dual to gravity states with deep throats. We saw that the deformation operator $D$ leads to the large energy gap (\ref{jjone}) changing to a very small energy gap due to band formation. Now we consider states in the CFT for which the  dual in the gravity theory has a short throat; in fact the energy gap in these short throats agrees perfectly with the energy gap in the CFT at the orbifold point. So the question is: why does the deformation operator $D$ not lead to band formation in this case?

Consider,  the state 
\be
|\psi_1\rangle=\left(\alpha_{A\dot A,-1}^{[1]}\bar\alpha_{B\dot B,-1}^{[1]}|0^+\rangle_R^{[1]}|\bar 0^+\rangle_R^{[1]}\right)\left(|0^+\rangle_R^{[1]}|\bar 0^+\rangle_R^{[1]}\right)^{N-1}
\label{sixpq}
\ee 
This differs from (\ref{initial}) in that the singly wound strings have a different ground state.

Consider the left moving part of the state; a similar situation holds for the right movers. The left moving part of the state is
\be\label{all +}
\left(\alpha^{[1]}_{A\dot A,-1}|0^+\rangle_R^{[1]}\right)\left( |0^+\rangle_R^{[1]}\right)^{N-1}
\ee
Note that each component string carries a  left moving  $SU(2)_L$ charge $1/2$.
Suppose the first component string carrying the $\alpha$ excitation joins with $k-1$ other strings to yield a $k$-wound component string. The $SU(2)_L$ charge contributed by the singly wound strings is  ${k\over 2}$. This charge must be carried by the $k$-wound component string. 

But the $SU(2)_L$ charge  is not carried by bosons; it must arise from fermions on this $k$-wound component string.  By the Pauli exclusion principle, we cannot put two fermions in the same state. The fermions do have two different flavors arising from the two different values of the charge $A$. If the total  $SU(2)_{L}$ charge is $j+\h$, then noting that each fermion carries $SU(2)_{L}$ charge $\h$, we find that the lowest energy state has the form (for integer $j$)
\be
d^{++[k]}_{-\frac{j}{k}}d^{+-[k]}_{-\frac{j}{k}}\dots d^{++[k]}_{-\frac{2}{k}}d^{+-[k]}_{-\frac{2}{k}}d^{++[k]}_{-\frac{1}{k}}d^{+-[k]}_{-\frac{1}{k}}d^{++[k]}_{0}d^{+-[k]}_{0}|0^-\rangle_R^{[k]}
\label{qlastsection}
\ee
This state has following dimension above the Ramond ground state $|0^-\rangle_R^{[k]}$
\be\label{dim k}
 h= \frac{j(j+1)}{k}
\ee
Setting $j+{1\over 2}={k\over 2}$ gives
\be
 h=\frac{k^2-1}{4k}
\ee
The initial excitation energy is carried by an operator $\alpha^{[1]}_{A\dot A,-1}$, which has dimension $1$. Thus if we have to maintain some approximate energy conservation during the evolution, then the twist sectors that we can reach are limited to the range
\be
k\sim 1
\ee

We now see that the initial excitation (\ref{sixpq}) cannot spread far along a lattice of points describing higher values of $k$. In an effective quantum mechanical model of the kind studied in the above sections, the energy (\ref{dim k}) yields a potential energy at each lattice site that rises with the value of $k$. Thus we do not have a situation where hopping can take place between sites of the same energy. Thus there  is no band formation. 

\section{Discussion}

In this paper we have proposed a resolution of a long-standing puzzle with the AdS/CFT duality map for the D1D5 system. At the orbifold point, we can consider two classes of Ramond ground states. Both are in the singly wound sector, but differ in the spins carried by the singly wound component strings. Since the component strings are singly wound, the energy gap at the orbifold point is $\Delta E=2$. But at in the supergravity regime the energy gap is $\Delta E=2$ for one class of states, and $\Delta E\ll 2$ for the other class. Thus the question is: how does the deformation operator $D$ in the CFT change the spectrum in this radical way when one moves off the orbifold point towards the supergravity point, and why thus change happens for one class of states and not the other.

All D1D5 Ramond ground states are protected by supersymmetry from any change in their energy; thus they remain Ramond ground states as we move off the orbifold point. At the supergravity point, they give the 2-charge extremal fuzzballs. We are however looking at the non-BPS excitations of such states which are caused by throwing a scalar quantum into the supergravity geometry; in the CFT at the orbifold point  this excitation is described by a left and a right bosonic excitation on one of the component strings. At the orbifold point this excitation in the CFT is an eigenstate of the Hamiltonian, with an energy $E=2$ that agrees with one set of supergarvity states (the ones with short throats) but not with the other set (the one with long throats). In the latter case the singly wound component strings carry no charge; thus they are able to join up into longer component strings. Thus the eigenstates at the orbifold point get linked by the deformation operator $D$ into a `band' with a small energy gap, and this situation does reflect the small energy gap found for the corresponding supergravity solutions which had a deep throat. Band formation was not allowed for the CFT states dual to short throats; in this case the singly wound component strings carried charge, and thus could not join into longer component strings at the same energy.  This again allowed an agreement of energy gaps between the CFT and the gravity descriptions.

We find the above phenomenon interesting because it illustrates how spacetime emerges from the CFT description. In the case where we have deep throats, we can say that a `lot of space' has emerged to form this throat. Where is this space in the CFT? Interactions link different CFT states together, creating an effective `lattice' on which excitations propagate. The length of this lattice is order $N\sim n_1n_5$, so the energy gap created is very small indeed.

Our analysis has been a preliminary one, and several further directions would be interesting to explore:

\b

(a) We have taken the initial excitation to have oscillators in the lowest allowed energy levels $\alpha_{-1}\bar\alpha_{-1}$. This energy formed a band for the case where the singly wound component strings did not carry any charge. If we take a higher energy excitation like $\alpha_{-2}\bar\alpha_{-2}$, then it will form its own band. A heuristic estimate indicates that at the supergravity point where the CFT coupling is $\lambda\gg 1$, these bands would overlap so that there is no energy gap between these bands. Such a situation would agree with the gravity dual, which does not exhibit any band of energies where quanta cannot be absorbed into the throat of the geometry. Exploring these higher energy excitations and band overlaps would therefore be interesting.

\b

(b) We have seen that the interaction $D$ leads to a 2-dimensional lattice with a slowly varying strength of the hopping term. It would be good to understand the exact band structure for such a lattice.

\b

(c) In our preliminary analysis we have assumed that transitions under the deformation operator $D$ take place from the initial state only to other states with the same energy. But since the evolution is a dynamical process, there can be `off-shell' processes where states with energies somewhat different from the initial energy are also involved. This increases the number of states that the wavefunction can flow to in our lattice of states. These off-shell effects are expected to become more important as we increase the coupling to go further away from the orbifold point.

\b

(d) We have considered the case where in the starting CFT state had all component strings with the same winding (unity). Now consider a generic Ramond ground state which has component strings with different windings $k_i$. When a string with a general winding $k_i$ joins up with a string with winding $k$, none of the  energy levels on the $k+k_i$ wound string may match the energy of the excitation that we had on the $k$-wound string. This means that the lattice of states on which we wish to diffuse does not have the same energy at each lattice site. A 1-d lattice with a random potential at each site leads to Anderson localization, which would prevent the band formation that we are looking for. As it turns out, we have a 2-dimensional lattice, and here if the random potential is sufficiently weak, then we will not have localization. 
These issues would be interesting to explore in more detail.

 \section*{Acknowledgements}

We would like to thank Marcel R. R. Hughes, Shaun Hampton, Yuri Kovchegov, Madhur Mehta for helpful discussions. The work of B.G. is supported by the ERC grant 787320-QBH Structure. The work of S.D.M. is supported in part by DOE grant DE-SC0011726.

\appendix

\section{Notations and conventions}\label{conventions}

We follow the notation in the appendix of \cite{hmz}. The indices $\alpha=(+,-)$ and $\bar \alpha=(+,-)$ correspond to the subgroups $SU(2)_L$ and $SU(2)_R$ arising from rotations on $S^3$. The indices $ A=(+,-)$ and $\dot A=(+,-)$ correspond to the subgroups $SU(2)_1$ and $SU(2)_2$ arising  from rotations in $T^4$. We use the convention
\be
\epsilon_{+-}=1, ~~~\epsilon^{+-}=-1
\ee

Consider the winding sector $(k_1,k_2,...,k_i,...)$
with the total winding $N=\sum_{i} k_i$. For the $i$th twisted set of copies  with winding number $k_{i}$,
we have following mode expansions on the cylinder.\footnote{Note that our convention of normalization has an extra factor of $1/ \sqrt{k_i}$ compared to the convention in \cite{gm2,gm3}.}
\bea\label{i string modes}
\alpha^{(i)}_{A \dot A,n}=\frac{1}{2\pi \sqrt{k_i}}\int_{\sigma=0}^{2\pi k_{i}}\p_{w}X^{(i)}_{A \dot A}(w)e^{nw}dw\nn
d^{\alpha A (i)}_{n}=\frac{1}{2\pi i \sqrt{k_i}}\int_{\sigma=0}^{2\pi k_{i}}\psi^{\alpha A(i)}(w)e^{nw}dw
\eea
In terms of $\alpha$ and $d$ modes, the $J$, $G$ and $L$ modes can be written as
\bea\label{i string modes J G L}
J^{a(i)}_m &=& {1\over 4 }\sum_{r}\epsilon_{AB}d^ {\gamma B(i)}_r\epsilon_{\alpha\gamma}(\sigma^{aT})^{\alpha}_{\beta}d^ {\beta A(i)}_{m-r},\qquad a=1,2,3\cr
J^{3(i)}_m &=&  - {1\over 2 }\sum_{r} d^ {+ +(i)}_{r}d^ {- -(i)}_{m-r} - {1\over 2 }\sum_{r}d^ {- +(i)}_r d^ {+ -(i)}_{m-r}\cr
J^{+(i)}_m&=&\sum_{r}d^ {+ +(i)}_rd^ {+ -(i)}_{m-r} ,\qquad J^{-(i)}_m=\sum_{r}d^ {--(i)}_rd^ {- +(i)}_{m-r}\cr
G^{\alpha(i)}_{\dot{A},r} &=& -i\sum_{n}d^ {\alpha A(i)}_{r-n} \alpha^{(i)}_{A\dot{A},n}\cr
L^{(i)}_m&=& -{1\over 2 }\sum_{n} \epsilon^{AB}\epsilon^{\dot A \dot B}\alpha^{(i)}_{A\dot{A},n}\alpha^{(i)}_{B\dot{B},m-n}- {1\over 2 }\sum_{r}(m-r+{1\over2})\epsilon_{\alpha\beta}\epsilon_{AB}d^ {\alpha A(i)}_r d^ {\beta B(i)}_{m-r}
\eea
Let $q$ be an integer. 
The mode numbers for $\alpha,L,J$ are $n=q/k_{i}$.
In the R sector, the mode numbers for $d$ and $G$ are $n=q/k_{i}$.
In the NS sector, the mode numbers for $d$ and $G$ are $n=(q+\frac{1}{2})/k_{i}$.
The modes (\ref{i string modes})  and (\ref{i string modes J G L}) satisfy the contracted large $\mathcal N=4$ superconformal algebra (\ref{app com a d}), (\ref{app com current a d}) and (\ref{app com currents}) with $c=6k_{i}$.

The commutation relations for the contracted large $\mathcal N=4$ superconformal algebra are
\bea\label{app com a d}
[\alpha_{A\dot{A},m},\alpha_{B\dot{B},n}] &=& -m\epsilon_{A B}\epsilon_{\dot A \dot{B}}\delta_{m+n,0}\cr
\{d^{\alpha A}_r , d^{\beta B}_s\}  &=&-\epsilon^{\alpha\beta}\epsilon^{AB}\delta_{r+s,0}
\eea
\bea\label{app com current a d}
[L_m,\alpha_{A\dot{A},n}] &=&-n\alpha_{A\dot{A},m+n} ~~~~~~~[L_m ,d^{\alpha A}_r] =-({m\over2}+r)d^{\alpha A}_{m+r}\cr
\lbrace G^{\alpha}_{\dot{A},r} ,  d^{\beta B}_{s} \rbrace&=&i\epsilon^{\alpha\beta}\epsilon^{AB}\alpha_{A\dot{A},r+s}~~~~~~~
[G^{\alpha}_{\dot{A},r} , \alpha_{B \dot{B},m}]=  -im\epsilon_{AB}\epsilon_{\dot{A}\dot{B}}d^{\alpha A}_{r+m}\cr
[J^a_m,d^{\alpha A}_r] &=&{1\over 2}(\sigma^{Ta})^{\alpha}_{\beta}d^{\beta A}_{m+r}
\eea
\bea\label{app com currents}
[L_m,L_n] &=& {c\over12}m(m^2-1)\delta_{m+n,0}+ (m-n)L_{m+n}\cr
[J^a_{m},J^b_{n}] &=&{c\over12}m\delta^{ab}\delta_{m+n,0} +  i\epsilon^{ab}_{\,\,\,\,c}J^c_{m+n}\cr
\lbrace G^{\alpha}_{\dot{A},r} , G^{\beta}_{\dot{B},s} \rbrace&=&  \epsilon_{\dot{A}\dot{B}}\bigg[\epsilon^{\alpha\beta}{c\over6}(r^2-{1\over4})\delta_{r+s,0}  + (\sigma^{aT})^{\alpha}_{\gamma}\epsilon^{\gamma\beta}(r-s)J^a_{r+s}  + \epsilon^{\alpha\beta}L_{r+s}  \bigg]\cr
[J^a_{m},G^{\alpha}_{\dot{A},r}] &=&{1\over2}(\sigma^{aT})^{\alpha}_{\beta} G^{\beta}_{\dot{A},m+r}\cr
[L_{m},J^a_n]&=& -nJ^a_{m+n}\cr
[L_{m},G^{\alpha}_{\dot{A},r}] &=& ({m\over2}  -r)G^{\alpha}_{\dot{A},m+r}
\eea
We define $J^{+}_n, J^-_n$ as
\bea
J^+_n &=& J^1_n + i J^2_n\cr
J^-_n&=& J^1_n - i J^2_n
\eea
From (\ref{app com current a d}), one can see that $d^{\alpha A}_{n}$ with $\alpha=+,-$ is a $SU(2)_L$ charge doublet. We have
\bea
[J^{+}_m,d^{+ A}_r] &=& 0,\qquad~~~~~ [J^{-}_m,d^{+ A}_r] ~=~ d^{-A}_{m+r}\cr
[J^{+}_m,d^{- A}_r] &=& d^{+A}_{m+r},\qquad [J^{-}_m,d^{- A}_r] ~=~ 0
\eea
From (\ref{app com currents}), one can see that $G^{\alpha}_{\dot{A},r}$  with $\alpha=+,-$ is also a $SU(2)_L$ charge doublet. We have
\bea
[J^{+}_{m},G^{+}_{\dot{A},r}]  &=& 0 ,\qquad\qquad ~~~[J^{-}_{m},G^{+}_{\dot{A},r}]  ~=~ G^{-}_{\dot{A},m+r}\cr
[J^{+}_{m},G^{-}_{\dot{A},r}]  &=&G^{+}_{\dot{A},m+r},\qquad ~[J^{-}_{m},G^{-}_{\dot{A},r}]  ~=~ 0 
\eea

\section{The effect of twist operator}\label{app twist}

In this appendix, we list all quantities defined in section \ref{sec twist}. See \cite{c4} for more details.
We define $a=e^{w_0/2}$.

\b

(i) Contractions: 
For bosons in the same copy 
\bea
C\left[\alpha^{[M]}_{A\dot A,-\frac{n_1}{M}}\,\alpha^{[M]}_{B\dot B,-\frac{n_2}{M}}\right]
&=&\frac{1}{M}\sum_{k\geq 0}^{n_2-1}{}^{-\frac{N}{M}n_1}C_{n_1+n_2-k}\,{}^{-\frac{N}{M}n_2}C_{k}\,(-n_2+k)\nn
&&~\times (-a)^{-(\frac{N}{M}+1)(n_1+n_2)}\,\epsilon_{A B}\,\epsilon_{\dot A \dot B}
\eea
For bosons in different copies
\bea
C\left[\alpha^{[M]}_{A\dot A,-\frac{n_1}{M}}\,\alpha^{[N]}_{B\dot B,-\frac{n_2}{N}}\right]
&=&\frac{1}{\sqrt{M\,N}}\sum_{p\geq 0}^{n_2-1}\,\sum_{k\geq 0}^{n_1-1}{}^{-\frac{M}{N}n_2}C_{p}\,{}^{-n_2+p}C_{n_1-k}\,{}^{-\frac{N}{M}n_1}C_{k}\,(-1)^p\,(-n_1+k)
\nn
&&~\times a^{-(\frac{M}{N}+1)n_2-(\frac{N}{M}+1)n_1}\,(-1)^{-(\frac{N}{M}+1)n_1-n_2+1}\,\epsilon_{A B}\,\epsilon_{\dot A \dot B}
\eea
For fermions in the same copy
\bea
C\left[d^{+A[M]}_{-\frac{n_1}{M}}\,d^{-B[M]}_{-\frac{n_2}{M}}\right]
&=&\Bigg(\frac{M+N}{M}\sum_{k\geq 0}^{n_2-1}{}^{-\frac{N}{M}n_1-1}C_{n_1+n_2-k-1}\,{}^{-\frac{N}{M}n_2}C_{k}\nn
&&~~~~~~~~+\sum_{k\geq 0}^{n_2-1}{}^{-\frac{N}{M}n_1-1}C_{n_1+n_2-k}\,{}^{-\frac{N}{M}n_2}C_{k}\Bigg)\nn
&&~~\times (-a)^{-(\frac{N}{M}+1)(n_1+n_2)}\,\epsilon^{AB}
\eea
For fermions in different copies
\bea
C\left[d^{+A[M]}_{-\frac{n_1}{M}}\,d^{-B[N]}_{-\frac{n_2}{N}}\right]
&=&\Bigg(\frac{M+N}{\sqrt{M\,N}}\sum_{p\geq 0}^{n_1-1}\,\sum_{k\geq 0}^{n_2-1}{}^{-\frac{N}{M}n_1-1}C_{p}\,{}^{-n_1+p}C_{n_2-k-1}\,{}^{-\frac{M}{N}n_2}C_{k}\,(-1)^p \nn
&&~~~~+\sqrt{\frac{M}{N}}\sum_{p\geq 0}^{n_1}\,\sum_{k\geq 0}^{n_2-1}{}^{-\frac{N}{M}n_1-1}C_{p}\,{}^{-n_1+p-1}C_{n_2-k-1}\,{}^{-\frac{M}{N}n_2}C_{k}\,(-1)^{1+p}\Bigg)
\nn
&&~~\times a^{-(\frac{N}{M}+1)n_1-(\frac{M}{N}+1)n_2}\,(-1)^{-(\frac{M}{N}+1)n_2-n_1}\,\epsilon^{B A}
\eea
By exchanging $M\leftrightarrow N$, one can get the contractions for modes with $[M]\leftrightarrow [N]$.

\b

(ii) Propagations: Define
\be
\mu_s=1-e^{2\pi i M s}
\ee
we have
\begin{equation}\label{f B}
\begin{split}
 f^{B(M)}_{rs} = & \begin{cases}
{\sqrt M \over \sqrt{M+N}} & r = s \\
{i\over 2\pi}a^{2(s-r)}
{\mu_s \over s-r}\,
{1\over  \sqrt{ M(M+N)}}\left( {(M+N)^{M+N}\over M^MN^N}\right)^{s-r}
{\Gamma[(M+N)r]\over \Gamma[Mr]\Gamma[Nr]}\,
{\Gamma[Ms]\Gamma[Ns]\over \Gamma[(M+N)s]}
\qquad & r \neq s 
\end{cases}\qquad\quad \cr
 f^{B(N)}_{rs}  = & \begin{cases}
{\sqrt N \over \sqrt{M+N}} & r = s \\
-{i\over 2\pi}a^{2(s-r)}
{\mu_s\over s-r}\,
{1\over  \sqrt{ N(M+N)}}\left( {(M+N)^{M+N}\over M^MN^N}\right)^{s-r}
{\Gamma[(M+N)r]\over \Gamma[Mr]\Gamma[Nr]}\,
 {\Gamma[Ms]\Gamma[Ns]\over \Gamma[(M+N)s]} \qquad & r \neq s
\end{cases}\qquad\quad
\end{split}
\end{equation}

\begin{equation}\label{f M}
\begin{split}
f_{rs}^{F(M)+}=&\begin{cases}
\frac{\sqrt{M}}{\sqrt{M+N}}&\qquad r=s\\
&\\
\frac{s}{r} f^{B(M)}_{rs}
 &\qquad r\neq s
\end{cases}
\hspace{2cm}
f_{rs}^{F(M)-}=
\begin{cases}\frac{\sqrt{M}}{\sqrt{M+N}}&\qquad r=s\\
&\\
f^{B(M)}_{rs}
& \qquad r\neq s
\end{cases}
\end{split}
\end{equation}

\begin{equation}\label{f N}
\begin{split}
f_{rs}^{F(N)+}=&\begin{cases}
\frac{\sqrt{N}}{\sqrt{M+N}}& \qquad r=s\\
&\\
\frac{s}{r}f^{B(N)}_{rs}
 & \qquad r\neq s
\end{cases}
\hspace{2cm}
f_{rs}^{F(N)-}=
\begin{cases}\frac{\sqrt{N}}{\sqrt{M+N}}&\qquad r=s\\
&\\
f^{B(N)}_{rs}
&\qquad r\neq s
\end{cases}
\end{split}
\end{equation}

\b

(iii) Pair creations: 

\bea
&&C_{MN}=\frac{M+N}{2MN}\nn
&&\gamma^{B}_{sr}=\frac{a^{2(s+r)}}{4\pi^2}\mu_{s}\mu_{r}\frac{1}{s+r}\frac{MN}{(M+N)^2}\left(\frac{(M+N)^{M+N}}{M^M N^N}\right)^{s+r}\frac{\Gamma[M s]\Gamma[N s]}{\Gamma[(M+N) s]}\frac{\Gamma[M r]\Gamma[N r]}{\Gamma[(M+N) r]}\nn
&&\gamma^{F}_{sr}=-s\,\gamma^{B}_{sr}
\eea

\end{document}